\documentclass[11pt,a4paper]{article}

\usepackage{jheppub}

\usepackage[english]{babel} 
\usepackage{amssymb}
\usepackage{graphicx}

\usepackage{amsmath}
\usepackage{commath}
\usepackage{amssymb}
\usepackage{comment}
\usepackage{braket}
\usepackage{ dsfont }
\usepackage{physics}
\usepackage{xcolor}
\usepackage[titletoc, title]{appendix}

\usepackage[utf8]{inputenc}

\usepackage{tikz}
\usetikzlibrary{positioning}
\usetikzlibrary{intersections}
\usetikzlibrary{fadings} 
\usetikzlibrary{arrows.meta}

\tikzfading[name=fade out,
inner color=transparent!0,
outer color=transparent!100]

\usetikzlibrary{decorations.pathmorphing}
\usetikzlibrary{decorations.pathreplacing,decorations.markings}

\definecolor{cherryblossompink}{rgb}{1.0, 0.72, 0.77}
\definecolor{lightblue}{rgb}{0.68, 0.85, 0.9}

\usepackage{subfigure}
\usetikzlibrary{backgrounds,automata}

\newcommand{\lp}{\left(}
\newcommand{\rp}{\right)}

\renewcommand{\figref}{figure~\ref}

\title{\boldmath No Page Curves for the de Sitter Horizon}

\author[a,c]{Joshua Kames-King,}
\author[b]{Evita M.H. Verheijden,}
\author[b]{and Erik P. Verlinde}

\affiliation[a]{Bethe Center for Theoretical Physics \& Physikalisches Institut der Universit\"at Bonn, \\
Nussallee 12, 53115 Bonn, Germany\\}
\affiliation[b]{Institute of Physics \&
Delta Institute for Theoretical Physics,\\ University of Amsterdam, Science Park 904, \\
1090 GL Amsterdam, The Netherlands\\}
\affiliation[c]{Kavli Institute for Theoretical Physics,\\
	University of California, Santa Barbara, CA93106, USA}

\emailAdd{jvakk@yahoo.com}
\emailAdd{e.m.h.verheijden@uva.nl}
\emailAdd{e.p.verlinde@uva.nl}

\abstract{We investigate the fine-grained entropy of the de Sitter cosmological horizon. Starting from three-dimensional pure de Sitter space, we consider a partial reduction approach, which supplies an auxiliary system acting as a heat bath both at $\mathcal{I}^+$ and inside the static patch. This allows us to study the time-dependent entropy of radiation collected for both observers in the out-of-equilibrium Unruh-de Sitter state, analogous to black hole evaporation for a cosmological horizon. 
Central to our analysis in the static patch is the identification of a weakly gravitating region close to the past cosmological horizon; this is suggestive of a relation between observables at future infinity and inside the static patch. 
We find that in principle, while the meta-observer at $\mathcal{I}^+$ naturally observes a pure state, the static patch observer requires the use of the island formula to reproduce a unitary Page curve. However, in practice, catastrophic backreaction occurs at the Page time, and neither observer will see unitary evaporation. 
}

\begin{document} 
\maketitle
\flushbottom

\baselineskip16pt

\vspace{0.5cm}
\section{Introduction}\label{sec:intro}
One of the greatest puzzles posed by our Universe---which we to good approximation believe to be described by de Sitter space---is a proper understanding of the cosmological horizon that surrounds any observer. The cosmological horizon of such a static observer exhibits thermodynamic properties similar to a black hole horizon \cite{GibbonsHawking}. One of the subtle obscurities is the entropy associated to the cosmological horizon, and in particular the fact that it appears to be finite. This seems to imply a finite-dimensional Hilbert space, which is in direct contradiction with the infinite-dimensional degrees of freedom of effective field theory on a de Sitter background \cite{Wittendesitter,troublewithds,ParikhErik,Erik2,higherspinds,Dong:2018cuv,Geng:2019bnn}. This discrepancy constitutes a significant problem, as not only the early Universe but also the current Universe at large scales is approximated by de Sitter space. While a complete microscopic understanding would require a full quantum gravity approach, here we follow a semi-classical approximation very much in the spirit of recent developments in a black hole context \cite{RyuTakayanagi,HubenyRT,EngelhardtQES,Faulkner:2013ana,Barrella:2013wja,Penington1,AlmheiriEMM,AlmheiriMMZ}. In practice this means that, while the exact microscopic state may be unknown, there is still a procedure to calculate the fine-grained entropy.

Until recently it was not known how unitary evolution of an evaporating black hole could be aligned with the apparently ever-increasing entropy of Hawking radiation. In a series of breakthroughs it was shown that the Quantum Extremal Surface (QES) of a non-gravitating region entangled with a gravitational system undergoes a phase transition at the Page time: the empty surface jumps to a surface just behind the horizon. This implies that the Hawking radiation follows the Page curve in accordance with unitarity \cite{Page1,Page2}. This result is now seen as an instantiation of a more general rule:\footnote{We are neglecting (potential) subtleties here about the use of the island rule in a gravitating system, see \cite{Raju1,Raju2}.}
\begin{equation}\label{eq:island}
	S_{\rm QG}[\text{Rad}] = \underset{I}{\text{min}}\Big\{ \underset{I}{\text{ext}} \Big[ S_{\text{SCG}}[\text{Rad} \cup I] + \frac{\text{Area}[\partial I]}{4G}\Big] \Big\}~.
\end{equation}
This so-called `island rule' tells us that the calculation of the fine-grained entropy must allow for the existence of disconnected regions or `islands'. To compute the entropy of Hawking radiation in quantum gravity (QG), we should include ``quantum extremal islands" in our semi-classical entropy calculation (SCG). These islands can minimize the entropy, e.g.\ an island just inside the black hole horizon will include Hawking partners of the radiation. The price to pay is the area of the island. Finally, one has to extremize and minimize over all possible islands. 

The island rule has been used to reproduce the Page curve for various black hole solutions \cite{AlmheiriOutside,Hollowood1,MyersEquil,Almheirihigherdimensions,Gautason:2020tmk,Anegawa:2020ezn,Hashimoto:2020cas,Hartman:2020swn,higherD2,Alishahiha:2020qza}. We would like to use these developments to learn more about the cosmological horizon. More specifically, we extend the procedure of \cite{Verheijden}, in which the evaporation of two-dimensional black holes in JT gravity on AdS$_2$ was studied from a three-dimensional point of view, to de Sitter space. In the original setup, the authors effectively divided the BTZ black hole into two parts. In one of these, they integrated out the angular coordinate, thereby reducing it to a black hole in AdS$_2$ JT gravity. In the other part, the holographic coordinate was integrated out, thereby obtaining the dual CFT (which took on the role of the `bath'). The evaporation of the 2D black hole effectively corresponds to changing the location of the dividing line. The entropy of a region in the bath can then be computed using geodesics in the three-dimensional BTZ geometry; this reproduces the Page curve for the bath entropy.

One might wonder if a similar approach could lead to new insights on the nature of the de Sitter entropy. A natural starting place would be to consider JT gravity in de Sitter \cite{Maloney,MaldacenaTuriaci,Blommaert:2020tht}. There has been some work on entanglement islands in a cosmological setup, see e.g.\ \cite{Sybesma,BalasubramanianDSislands,CosmologyIslands,Geng:2021wcq}. In particular, \cite{Aalsma:2021bit} provides a complementary perspective to the approach we will take, which we outline below.

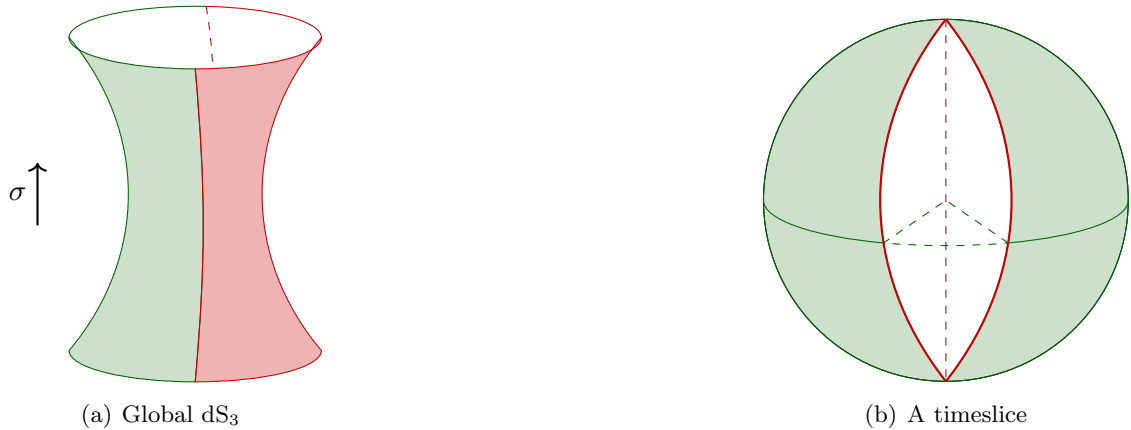
\begin{figure}
\centering
\subfigure[{Global dS$_3$}\label{fig:Setup1}]{
\begin{tikzpicture}[scale=0.83]

    \draw[thick, ->] (-2.5, 2) --++(90:1) node[pos=.5, left] {$\sigma$};
    \filldraw[fill=green!40!black, draw=green!40!black, fill opacity=0.2] (0,-0.5)  arc[start angle=270,end angle=180, x radius=2, y radius=0.5] to[bend right=40] (-2,5) arc[start angle=180, end angle=270, x radius=2cm, y radius=0.5cm] to[bend left=5] (0,-0.5);
    \draw[green!40!black] (-2,5) arc[start angle=180, end angle=85, x radius=2cm, y radius=0.5cm] coordinate (a);
    \draw[red!80!black] (a) arc[start angle=85, end angle=0, x radius=2cm, y radius=0.5cm];
    \path (0,4.5) arc[start angle=270, end angle=278, x radius=2cm, y radius=0.5cm] coordinate (b);
    \draw[dashed, red!80!black] (a) to[bend left=3] (b);
    \filldraw[draw=red!80!black, fill=red!80!black, fill opacity=0.3] (0,4.5) arc[start angle=270, end angle=360, x radius=2cm, y radius=0.5cm] to[bend right=40] (2,0) arc[start angle=0, end angle=-90, x radius=2cm, y radius=0.5cm] to[bend right=5] (0,4.5);
\end{tikzpicture}
}
\hfill \subfigure[{A timeslice}\label{fig:Setup2}]{
\begin{tikzpicture}[scale=1.2]
    \filldraw[draw=transparent, fill=green!40!black, fill opacity=0.2] (0,2) to[bend right=38] (0,-2) arc[start angle=270, end angle=90, radius=2cm];
    \filldraw[draw=transparent, fill=green!40!black, fill opacity=0.2] (0,2) to[bend left=38] (0,-2) arc[start angle=-90, end angle=90, radius=2cm];
    \draw[name path = {leftline}, thick, red!80!black] (0,2) to[bend right=38] (0,-2);
    \draw[name path = {rightline}, thick, red!80!black] (0,2) to[bend left=38] (0,-2);
    \path[name path = {equator}] ellipse (2cm and 0.5cm);
    \draw[green!40!black, name intersections={of=equator and leftline, by={up1,down1}}] (down1) arc[start angle=250, end angle=180, x radius=2cm, y radius=0.5cm];
    \draw[green!40!black, name intersections={of=equator and rightline, by={up2,down2}}] (down2) arc[start angle=290, end angle=360, x radius=2cm, y radius=0.5cm];
    \draw[dashed, green!40!black] (down2)--(0,0) -- (down1) arc[start angle=250, end angle=290, x radius=2cm, y radius=0.5cm];
    \draw[dashed, red!80!black] (0,2)--(0,-2);
    \draw[green!40!black] (0,0) circle (2cm);
\end{tikzpicture}
}
\caption{Global de Sitter is the surface of the hyperboloid (a). Time flows upwards; one angular coordinate is suppressed, such that each time-slice is actually a two-sphere (b). We split the spacetime into two, reducing over the red part to get JT gravity  on dS$_2$. The green part is the remainder of dS$_3$, which takes on the role of the bath.}
\label{fig:Setup}
\end{figure}

\paragraph{Our approach.} We will start from pure (empty) dS$_3$ and perform a similar trick as explained above. A partial dimensional reduction of  dS$_3$ along the angular direction $\varphi$ divides the three-dimensional spacetime into two. Up to some value of the angular coordinate the system is described by dynamical gravity: JT gravity on dS$_2$. The remainder of the three-dimensional spacetime will take on the role of the thermal `bath' for the radiation of the cosmological horizon. To be precise, in our two-dimensional set-up we will identify two regions where gravity is weakly coupled; these regions may be considered non-dynamical and are thus good regions to collect radiation. We will then compute the fine-grained entropy of radiation collected in these regions by embedding them in the three-dimensional geometry. In this sense, we will refer to the remainder of the three-dimensional spacetime as the non-gravitating `bath'. The full (global) setup is depicted in \figref{fig:Setup}.

\begin{figure}
\centering
\begin{tikzpicture}[scale=1.2]
 
     \fill[green!60!black, opacity=0.6, path fading=fade out] (0,2) -- (0,3) arc[start angle=90, end angle=340, radius=3cm] --++(160:1) arc[start angle=340, end angle=90, radius=2cm];
    \draw[black, thick] (0,2) arc[start angle=90, end angle=340, radius=2cm] coordinate (b);
    \draw[black, thick] (0,2) arc[start angle=90, end angle=-20, radius=2cm];
    \draw[gray, dashed] (0,1.6) arc[start angle=90, end angle=340, radius=1.6cm];
    \node at (2.5,-2) {\small{cosmological horizon}}; 
    \draw[red!80!black, thick] (0,0) -- (0,2);
    \draw[red!80!black, thick] (0,0) --++(-20:2);
    \draw[red!80!black, thick, path fading=north] (0,2)--(0,3);
    \draw[red!80!black, thick, path fading=south] (b)--++(340:1);
    \filldraw[draw=transparent, fill=green!60!black, opacity=0.2] (0,2) arc[start angle=90, end angle=340, radius=2cm] coordinate (C) -- (0,0) -- cycle;
    \filldraw[draw=red!80!black, fill=red!80!black, fill opacity=0.2, draw opacity=0.2] (0,2) arc[start angle=90, end angle=-20, radius=2cm] --(0,0)--cycle;

    \draw[<->, gray] (0,0.5) arc[start angle=85, end angle=-20, radius=0.5cm] node[pos=.5, right] {$\alpha(t)$};
    
    \draw[orange, decorate, decoration={snake, segment length=0.2cm, amplitude=0.4mm}, ->] (0, 1.8) arc[start angle=90, end angle=110, radius=1.8cm];
    \path (0,1.8) arc[start angle=90, end angle=125, radius=1.8cm] coordinate (rad2);
    \draw[orange, decorate, decoration={snake, segment length=0.2cm, amplitude=0.4mm}, ->] (rad2) arc[start angle=125, end angle=145, radius=1.8cm];
  
    \path (b) --++(160:0.20) coordinate (r1) --++(160:0.30) coordinate (r2) --++(160:0.25) coordinate (r3) --++(160:0.25) coordinate (r4) --++(160:0.25) coordinate (r5);
    \draw[orange, decorate, decoration={snake, segment length=0.2cm, amplitude=0.4mm}, ->] (r1) arc[start angle=340, end angle=320, radius=1.8cm];
    \path (r1) arc[start angle=340, end angle=305, radius=1.8cm] coordinate (rad3);
    \draw[orange, decorate, decoration={snake, segment length=0.2cm, amplitude=0.4mm}, ->] (rad3) arc[start angle=305, end angle=285, radius=1.8cm];
    \end{tikzpicture}
\caption{ A timeslice of the three-dimensional static patch in our partial reduction approach. In the dual description, radiation moves along the stretched horizon. This has the effect of shifting the dividing line between bath and JT gravity. The entanglement between these two systems is a function of the angle $\varphi$ (parametrised by $\alpha$), which dynamically decreases during the evaporation process. We also indicated the part of the 3D geometry that has been reduced over, to clarify the entanglement structure.}
\label{fig:staticpatchevaporationpicture}
\end{figure}
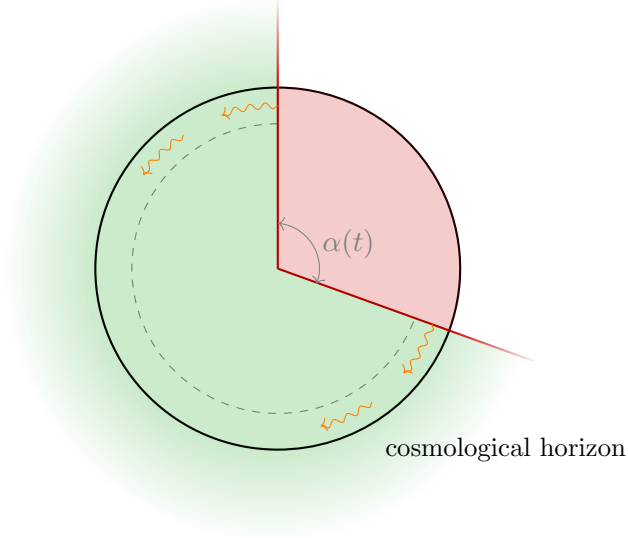

Motivated by recent results in the context of evaporating black holes, we will consider an out-of-equilibrium thermal state corresponding to the evaporation of the cosmological horizon. As shown in \cite{backreactionFuture}, this so called Unruh-de Sitter state amounts to demanding a positive net incoming energy flux on the static patch, breaking the isometries preserved in the standard Bunch-Davies vacuum. The Schwarzian dynamics of $\mathcal{I}^+$, established in \cite{Maloney,MaldacenaTuriaci}, allow for the calculation of the backreaction of the assumed matter configuration on the boundary dilaton, which becomes a function of the single boundary variable $u$ at future infinity.\footnote{Although $u$ appears as a spacelike coordinate at future infinity, we will often denote a function of this variable as `time-dependent'. We do so because we take the results of section \ref{sec:staticpatch} to mean that in the static patch functions of $u$ become functions of the proper static patch time.} From the three-dimensional perspective the renormalised boundary dilaton corresponds to the angle of the dimensional reduction, $\Phi_r \sim 2\pi\alpha$, such that the backreaction of the Unruh-de Sitter state imbues the full three-dimensional setup with a dependence on $u$. 
In the partial reduction the entropy of the cosmological horizon is
\begin{equation}\label{eq:introentropy}
     S_{\text{dS},\alpha}=\frac{\pi \alpha \ell}{2 G^{(3)}}\,,
\end{equation}
where $\alpha$ is the parameter determining the reduction angle. From \eqref{eq:introentropy} we see that a dynamical boundary dilaton $\Phi_r (u) \sim \alpha (u)$ not only amounts to dynamical evolution of the dividing line between the thermal bath and gravity, but also to a decreasing entropy. As depicted in \figref{fig:futureinfinityevaporationpicture} and \figref{fig:staticpatchevaporationpicture} of section \ref{sec:entropies}, in the three-dimensional picture we think of this dynamical change as the evaporation of the radiation into the bath. 

Note that in \cite{Verheijden}, it was the mass of the BTZ black hole that became time-dependent, and consequently the entropy; however, empty de Sitter only exhibits a single (fixed) length scale. We can still introduce a time-dependent entropy if we allow for time-dependence in $\alpha$ and hence consider $(\alpha\,\ell)$ as an effective time-dependent de Sitter length.

While the behaviour of $\alpha(u)$ can indeed be determined at the future boundary, we will also find that we can recover the same behaviour by use of an explicit bulk solution at the cosmological horizon in the static patch. This allows us to address an important subtlety that arises for the de Sitter case: we can make a choice of observer. Whereas the static observer is surrounded by a cosmological horizon and as such experiences a thermal bath, we can also define a `meta-observer' at future infinity, who can observe the wavefunction of the universe as they have access to distances larger than the Hubble scale \cite{PhysRevD.28.2960,Wittendesitter}. From a cosmological perspective we can (approximately) be described as a static observer currently entering a new de Sitter phase. However, we may also be considered meta-observers with respect to our inflationary past \cite{Danielsson:2002qh}.

As our construction creates a thermal bath in both the static patch and at future infinity of the two-dimensional de Sitter space, we can perform calculations for both observers. These are complementary views and we give results for the entropy of the collected radiation with respect to both. As a second subtlety, we must take into account the lifetime of the backreacted solution. In a semi-classical setting, various arguments have been made about the lifetime of de Sitter. We will consider how these approaches bound our results. Furthermore, as we are considering an out-of-equilibrium state, we should expect the lifetime to be drastically reduced and even a singularity to arise \cite{Aalsma:2021bit}. These considerations naturally will reduce the domain of validity of the entropy computations. \\

\noindent This paper is organised as follows. In section \ref{sec:reduction} we discuss how to obtain JT gravity on two-dimensional de Sitter from a three-dimensional Einstein-Hilbert action in de Sitter. We discuss the two-dimensional bulk equations of motion, and then comment on the boundary action and the dilaton at future infinity. In section \ref{sec:Adding matter:adynamicalboundarydilaton} we introduce dynamics by considering the effect of adding matter to our configuration. Specifying the Unruh-de Sitter state, we find a dynamical boundary dilaton $\Phi_r (u)$. This allows us to estimate the lifetime of our setup, and we comment on the timescales relevant to our problem. We also calculate the backreacted bulk dilaton in section \ref{sec:staticpatch}. We discover that the gravitational coupling becomes weak at the past horizon. Remarkably, the explicit backreacted bulk dilaton exhibits the same behaviour close to the past cosmological horizon in terms of the static patch time $t$ as the boundary dilaton does in terms of $u$. In section \ref{sec:entropies} we move on to the calculation of the entanglement entropy of the radiation as a function of our boundary time $u$ and of the static patch time $t$. We explain why our setup naturally supplies an auxiliary system at both $\mathcal{I}^{+}$ and in the static patch, and we compute the fine-grained entropy in both regions. While for $\mathcal{I}^{+}$ we recover unitary behaviour without the use of an island, the static patch requires a more involved argument and, implicitly, the existence of an island. This agrees with intuition due to the different locations of these regions with respect to the cosmological horizon. The meta-observer is in causal contact with behind the horizon degrees of freedom, whereas the static patch observer represents a thermal observer. However, the aforementioned finite lifetime of the Unruh de Sitter state corresponds to the occurrence of a trapped region at the Page time. Therefore information recovery does not seem possible for the evaporating de Sitter horizon. 

\flushbottom

\section{\texorpdfstring{JT gravity on dS$_2$ from dS$_3$}{JT dS2 gravity from dS3}} \label{sec:reduction}
 As outlined in the introduction, we start with three-dimensional gravity on pure de Sitter space. The first step is to perform a partial dimensional reduction on the spherical coordinate $\varphi$. This means we consider the upper value of the spherical integration of $\varphi$ to be given in terms of a new parameter $\alpha \in (0,1]$. As we will see below, $\alpha$ is closely related to the dilaton in two dimensions.

\subsection{Dimensional reduction from 3D Einstein to JT gravity}
Our starting point is given by the three-dimensional action
\begin{equation}\label{eq:threedimensionalaction}
    S = \frac{1}{16\pi G^{(3)}} \int d^3 x \sqrt{-g^{(3)}} \left( R^{(3)} - \frac{2}{\ell^2} \right) - \frac{1}{8\pi G^{(3)}} \int d^2 x \sqrt{-h^{(3)}} \left(K^{(3)}-1\right)\,,
\end{equation}
where the last term is the Gibbons-Hawking boundary term. Here $K^{(3)}$ plays an important role as it will furnish the Schwarzian boundary action at future infinity $\mathcal{I}^+$. The Einstein equations give $R^{(3)}=\frac{6}{\ell^2}$. 

\begin{figure}
\centering
\begin{tikzpicture}
	\pgfmathsetmacro\myunit{4}
	\draw	(0,0)			coordinate (a)
		--++(90:\myunit)	coordinate (b)
							node[pos=.5, above, sloped] {\text{North Pole}}
		--++(0:\myunit)		coordinate (c)
							node[pos=.5, above] {$\mathcal{I}^+$}
		--++(-90:\myunit)	coordinate (d)	
							node[pos=.5, sloped, above] {\text{South Pole}}
		--cycle 			node[pos=.5, below] {$\mathcal{I}^-$};
 	\draw[dashed, name path = hor1] (a) --++ (c);
 	\draw[dashed, name path = hor2] (b) -- (d);
 	\draw[dashed, fill=orange!50!white, fill opacity=0.5, name intersections={of=hor1 and hor2, by={int1}}] (d) -- (int1) -- (c);
 	\draw[dashed, fill=lightblue, fill opacity=0.5] (b) -- (int1) -- (c);
 	\draw (a) -- (b);
 	\draw (c) -- (d);
 	\draw (b) -- (c);
 	\draw[decorate, decoration={snake, segment length=1cm, pre=lineto, pre length=0.3cm, post = lineto, post length=0.1cm}] (0,4) to[bend right=15] (4,4);
 	\node at (2,3.4) {$x(u)$};
\end{tikzpicture}
\caption{The Penrose diagram of three- and two-dimensional pure de Sitter. For dS$_3$ each point represents a circle. The static patch for an observer at the south pole is indicated in orange; the dashed lines are the horizons. The Milne (future) patch is indicated in blue. For dS$_2$ we will make use of the fluctuating boundary geometry at $\mathcal{I}^{+}$ described by a Schwarzian action. $\mathcal{I}^{-}$ does not play a role in our considerations as we consider a quantum state that is singular at the past horizon.} \label{fig:Penrosediagram}
\end{figure}
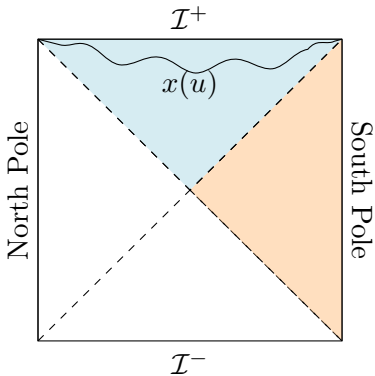

We collect different coordinate systems for de Sitter space in appendix \ref{app:coordinates}. Here, we single out two important systems we will use: global and static coordinates. In global conformal coordinates, three-dimensional de Sitter space is given by:
\begin{align}\label{eq:threedimensionalglobal}
    ds_3^2 = \frac{\ell^2}{\cos^2\sigma} \left(-\dd \sigma^2 + \dd \theta^2 + \sin^2\theta \dd \varphi^2\right)\,,
\end{align}
where $\sigma \in (-\frac{\pi}{2},\frac{\pi}{2})$, $\theta \in [0,\pi]$ and $\varphi \in [0, 2\pi)$. 
The corresponding Penrose diagram is a square, see \figref{fig:Penrosediagram}. No single observer can access the full geometry and there is no global timelike Killing vector.

The so-called static patch is the region accessible to a single observer living on one of the poles of the $S^2$. For this region, the $SO(1,3)$ isometry group gives rise to a manifest $t$-translation, such that we arrive at a time-independent metric, given by:
\begin{equation}\label{eq:3dstaticpatch}
        ds_3^2 = - \left(1-\frac{r^2}{\ell^2}\right) \dd t^2 + \left(1-\frac{r^2}{\ell^2}\right)^{-1} \dd r^2 + r^2 \dd \varphi^2\,,
    \end{equation}
with $r \in [0,\ell]$.
Note that the same angle $\varphi$ appears in both \eqref{eq:threedimensionalglobal} and \eqref{eq:3dstaticpatch}. The static coordinates cover only the right (orange) diamond of \figref{fig:Penrosediagram}. The null surface at $r=\ell$ surrounding the observer at all times is known as the cosmological horizon. The temperature associated to this horizon is
    \begin{equation}\label{eq:temperature}
        T_{\text{dS}}=\frac{1}{2 \pi \ell}\,,
    \end{equation}
which is a fixed quantity. The corresponding Gibbons-Hawking entropy is given by
    \begin{equation}\label{eq:dsentropy}
        S_{\text{dS}}=\frac{\pi \ell}{2 G^{(3)}}\,.
    \end{equation}
As outlined in the introduction, we consider a partial reduction ansatz by considering an upper value of the spherical coordinate $\varphi$ in the spherical integration to be given as $ 2\pi \alpha$, $\alpha \in (0,1]$, which then means that the Gibbons-Hawking entropy is given as \eqref{eq:introentropy}, which we repeat here for convenience:
     \begin{align}\label{eq:ourentropy}
        S_{\text{dS},\alpha}&=\frac{\pi \alpha \ell}{2 G^{(3)}} =\frac{\pi \alpha }{2 G^{(2)}}\,,
    \end{align}
where we identified $G^{(3)}=\ell G^{(2)}$ and we will from now on denote $G^{(2)}$ simply as $G$. We are interested in considering the evolution of \eqref{eq:ourentropy} by allowing for time-dependence in $\alpha$. We will see below that the dilaton of the JT theory we acquire in two dimensions is intimately linked to this angle $\alpha$. From the two-dimensional perspective, any backreaction is captured by the dilaton. Hence, it is clear that setting a two-dimensional Unruh-de Sitter state will not only lead to a time-dependent dilaton solution, but also from the higher-dimensional perspective lead to a dynamical change of \eqref{eq:ourentropy} as desired.

Let us now turn to the details of the dimensional reduction. We assume an ansatz of the form 
\begin{equation} \label{eq:dSreducAnsatz}
        ds_3^2 = g^{(2)}_{ij} \dd x^i \dd x^j + \phi^2(x^i) \ell^2 \dd \varphi^2\,.
    \end{equation}
Under the assumption of an asymptotic boundary, by use of the following identities 
 \begin{equation}
 \begin{aligned}
   R^{(3)}&=R^{(2)}-\frac{2}{\phi} \Box^{(2)}\phi \,, \\ 
   K^{(3)}&=K^{(2)}+\frac{1}{\phi} n^{ \mu}\nabla_{\mu}^{(2)} \phi~,
 \end{aligned}
 \end{equation}
and by integrating the spherical coordinate as outlined below \eqref{eq:dsentropy}, we find
    \begin{equation}\label{eq:JTaction1}
        S = \frac{2\pi \alpha \ell}{16\pi G^{(3)}} \int d^2 x \sqrt{-g^{(2)}} \phi \left(R^{(2)} - \frac{2}{\ell^2} \right) - \frac{2\pi \alpha \ell}{8\pi G^{(3)}} \int d  x \sqrt{-h^{(2)}} \phi_b \left(K^{(2)}-1\right) \,.
    \end{equation}
We have implicitly identified $\ell_3 = \ell_2 \equiv \ell$. Using again $G^{(3)}=\ell G$, we conclude 
    \begin{equation} \label{eq:bulkdilatondimred}
        \Phi= 2\pi \alpha \phi\,,
    \end{equation}
such that we arrive at the JT gravity action
    \begin{equation}\label{eq:JTaction2}
       S = \frac{1}{16\pi G} \int d^2 x \sqrt{-g^{(2)}}  \Phi \left(R^{(2)} - \frac{2}{\ell^2} \right) - \frac{1}{8\pi G} \int d  x \sqrt{-h^{(2)}} \Phi_b \left(K^{(2)}-1\right) \,. \end{equation}
Here $\Phi_b$ denotes the boundary value of $\Phi$. In the dimensionally reduced language of \eqref{eq:JTaction2} we recover the global metric \eqref{eq:threedimensionalglobal} by 
   \begin{equation} \label{eq:dilatonglobal}
        ds_2^2 = \frac{\ell^2}{\cos^2\sigma} \left(-\, \dd \sigma^2 + \dd \theta^2 \right)~, \qquad \Phi = 2\pi \alpha \frac{\sin\theta}{\cos\sigma}~.
    \end{equation}
The extrinsic curvature $K^{(2)}$ plays a pivotal role in our approach, but we will postpone our discussion of it to section \ref{sec:Boundaryaction}. First, we will expand on the bulk dynamics in section \ref{sec:2dbulkdynamics}.  Before we do so, we wish to point out that in \eqref{eq:JTaction2} we do not recover the Gauss-Bonnet term ordinarily used in JT gravity. This term, in an AdS context proportional to the ground state entropy of an extremal, higher-dimensional black hole, usually allows for negative values of the dilaton $\Phi$ while still maintaining positive values for the total entropy $\Phi_0 + \Phi$. In that case, the Penrose diagram of dS$_2$ `doubles' and allows for two horizons located at $r = \pm \ell$. The second horizon is often interpreted as a black hole horizon, and this geometry then serves as a lower-dimensional toy model for Schwarzschild-de Sitter. We are instead interested in studying `pure' de Sitter, and hence will stick to \figref{fig:Penrosediagram} also for the two-dimensional model, as we interpret this as inherited from the three-dimensional de Sitter spacetime. This agrees with the absence of $\Phi_0$ in \eqref{eq:JTaction2}.

\subsection{Two-dimensional bulk dynamics} \label{sec:2dbulkdynamics}
To study the two-dimensional bulk dynamics, it is convenient to switch to conformal gauge and employ general lightcone coordinates $(x^{+},x^{-})$:
\begin{equation}\label{eq:generallightconecoordinates}
    ds_2^2=-e^{2 \omega(x^{+},x^{-})}\dd x^{+} \dd x^{-}\,.
\end{equation}
In these coordinates the bulk equations of motion amount to \cite{AlmheiriPolchinski,BalasubramanianDSislands}
\begin{align}\label{eq:BulkEoms}
    \partial_{+}\partial_{-}\omega&=\frac{1}{4 \ell^2}e^{2 \omega} \nonumber\,,\\
    -\partial_{\pm}^2\Phi + 2 \partial_{\pm}\omega \partial_{\pm}\Phi&=8\pi G \langle T_{x^\pm x^\pm} \rangle \,,\\
    2\partial_{-}\partial_{+}\Phi-\frac{1}{\ell^2} e^{2 \omega}\Phi&=16 \pi G \langle T_{x^+ x^-} \rangle \nonumber\,.
\end{align}
 Let us now comment on different solutions to \eqref{eq:BulkEoms} in vacuum. As shown in appendix \ref{app:coordinates}, we can introduce null coordinates $\sigma^{\pm}$ such that the static patch metric \eqref{eq:3dstaticpatch} is given by
 \begin{equation}\label{eq:staticpatchnullcoordinates}
    e^{2 \omega(\sigma^{+},\sigma^{-})}=\frac{1}{\cosh^2\left(\frac{\sigma^+ - \sigma^-}{2 \ell}\right)}\,,\qquad \Phi=2\pi \alpha \frac{1}{\tanh\left(\frac{\sigma^{+}+\sigma^{-}}{2 \ell}\right)}\,.
 \end{equation}
 Again, these coordinates are restricted to the south pole wedge. We can also define Kruskal coordinates which cover the entire Penrose diagram, see \figref{fig:Penrosediagram}. As shown in appendix \ref{app:coordinates} this amounts to  \begin{equation}\label{eq:KruskalWatse}
    e^{2 \omega(x^{+},x^{-})}=\frac{4\ell^4}{(\ell^2-x^+ x^-)^2}\,,\qquad \Phi=2\pi \alpha\frac{\left(\ell^2+x^+ x^-\right)}{\left(\ell^2-x^+ x^-\right)}\,.
\end{equation}
The coordinate transformation that relates the Kruskal coordinates $(x^+,x^-)$ to the static coordinates $(\sigma^+,\sigma^-)$ is
    \begin{equation} \label{eq:Kruskaltostatic}
        x^{\pm}=\pm \ell e^{\pm \sigma^\pm/\ell}~,
    \end{equation}
which illustrates the different roles these coordinate systems play for us. The transformation is the same relationship as between Rindler and Minkowski coordinates, such that indeed the coordinate systems \eqref{eq:staticpatchnullcoordinates} and \eqref{eq:KruskalWatse} define different vacua.

In our approach we also care about the boundary dynamics of different solutions to \eqref{eq:BulkEoms}. When considering the desired non-equilibrium state, we should be able to see at future infinity $\mathcal{I}^{+}$ that the entropy \eqref{eq:ourentropy} has now become dynamical. The static patch \eqref{eq:staticpatchnullcoordinates} is not connected to the boundary, but it is connected via analytic continuation to the so-called Milne patch in the expanding region \cite{MaldacenaTuriaci}, see \eqref{eq:analyticcontinuation}. The Milne solution is 
\begin{equation}\label{eq:Milne}
    e^{2 \omega(y^{+},y^{-})}=\frac{1}{\sinh^2\left({\frac{y^{+}+y^{-}}{2\ell}}\right)}~, \qquad \Phi= 2\pi \alpha \frac{1}{\tanh \left(\frac{y^+-y^-}{2 \ell}\right)}~.
\end{equation}
In terms of the coordinates $(\tau, \chi)$ used in \eqref{eq:analyticcontinuation}, the lightcone coordinates $y^\pm$ used in \eqref{eq:Milne} are
    \begin{equation}
        y^\pm = \tilde{\tau} \pm \chi~, \qquad d\tau = \frac{d\tilde{\tau}}{\sinh{\tilde{\frac{\tau}{\ell}}}}~.
    \end{equation}
As we will see in the next section, this geometry is of importance for our purposes as it captures the evolution of the entropy \eqref{eq:ourentropy} at future infinity.

\subsection{Boundary action and renormalised dilaton}\label{sec:Boundaryaction}
We now turn to the boundary dynamics at $\mathcal{I}^{+}$ and the extrinsic curvature term of \eqref{eq:JTaction2}. In most of this section, we will use planar coordinates in order to make the analogy to \cite{MaldacenaNAdS} more apparent:
\begin{align}\label{eq:planarcoordinates}
    ds_2^2=\frac{\ell^2(-\dd \eta^2 + \dd x^2)}{\eta^2}\,,\;\;\;\;\;\Phi= - 2\pi \alpha \frac{x}{\eta}\,.
\end{align}
Note that $\eta \leq 0$ and $x \geq 0$, with $\mathcal{I}^+$ located at $\eta = 0$, such that the dilaton is correctly positive. 
As $\mathcal{I}^{+}$ is a conformal boundary, we would like to cut off the space along a boundary curve $\left(\eta(u), x(u) \right)$. It is usually conjectured that the complete gravitational theory can be described by a quantum mechanical system at the conformal boundary; then $u$ would correspond to the coordinate of this quantum mechanical boundary theory. It will play a special role in our setup as our results with respect to the entropy at future infinity are phrased in terms of this parameter. 
Following \cite{MaldacenaNAdS,MaldacenaTuriaci} we set the following two boundary conditions
\begin{equation}\label{eq:JTboundaryconditions}
    g_{u u}=\frac{\ell^2}{\epsilon^2}\,,\;\;\;\;\; \Phi_b=\frac{\Phi_r}{\epsilon}\,.
\end{equation}
For the partial reduction solutions we are considering, $\Phi_r$ generally takes on the form
\begin{equation}\label{eq:dilatonrenormalisedgeneral}
    \Phi_r=2 \pi \ell \alpha\,.
\end{equation}
Solving \eqref{eq:JTboundaryconditions} we get
\begin{equation}
    K^{(2)}=\frac{1}{\ell}-\frac{\epsilon^2}{\ell} \{x(u),u \}\,,
\end{equation}
such that the action \eqref{eq:JTaction2} reduces to the effective boundary term
\begin{equation}\label{eq:Schwarzianaction}
    S_{\rm GH}=\frac{1}{8\pi G}\int \dd u\, \Phi_r \{x(u),u \}\,.
\end{equation}
We can interpret \eqref{eq:Schwarzianaction} along the lines of \cite{MaldacenaNAdS}. The future boundary exhibits an asymptotic symmetry of reparametrisations of the coordinate $x(u)$, which may be understood as the gravitational degree of freedom of this two-dimensional system. By introducing the JT action, we explicitly break the symmetry and \eqref{eq:Schwarzianaction} may be considered the action of this `boundary graviton'.  
Variation of \eqref{eq:Schwarzianaction} with respect to the boundary mode $x(u)$ amounts to
\begin{equation}\label{eq:Schwarzianeom}
    -\frac{1}{8 \pi G}\left(\Phi_r \{x(u),u \}'+ 2 \Phi_r' \{x(u),u \}+\Phi_r''' \right)=0\,,
\end{equation}
where $'$ denotes derivation with respect to $u$.
We will ultimately be interested in a dynamical (renormalised) boundary dilaton $\Phi_r$. To fully understand the background solutions, let us first consider constant $\Phi_r$. Then \eqref{eq:Schwarzianeom} reduces to
\begin{equation}\label{eq:schwarzianeomtimeindependent}
    \{x(u),u \}'=0\,.
\end{equation}
The associated conserved charge is given by \cite{MaldacenaNAdS,HermanBackreaction,MaldacenaTuriaci}
\begin{equation}\label{eq:admquantity}
    K=-\frac{\Phi_r}{8 \pi G}\{x,u \}\,.
\end{equation}
One possible solution of \eqref{eq:schwarzianeomtimeindependent} is given by
\begin{align}\label{eq:thermalx}
    x(u)&= 2 \ell \tanh{\frac{u}{2 \ell}} =\frac{\beta_{\text{dS}}}{\pi} \tanh{\frac{\pi u }{\beta_{\text{dS}}}}\,,
\end{align}
which just corresponds to the Milne solution \eqref{eq:Milne}. 
Note that with \eqref{eq:dilatonrenormalisedgeneral} and \eqref{eq:thermalx}, \eqref{eq:admquantity} amounts to
\begin{equation}\label{eq:AdMtoentropy}
    K=\frac{S_{\text{dS},\alpha}}{2 \beta_{\text{dS}}}\,.
\end{equation}
Here we can see why the solution \eqref{eq:thermalx} (i.e., the metric \eqref{eq:Milne}) is of special importance to us. The ADM quantity \eqref{eq:AdMtoentropy} is linked to the entropy \eqref{eq:ourentropy} of the static observer, such that demanding dynamical behaviour of \eqref{eq:admquantity} at $\mathcal{I}^+$ conforms with the desired change in entropy. Hence, the sensitivity of the static patch entropy to the state of our matter fields in static coordinates will in section \ref{sec:Unruhstate} translate to the sensitivity of \eqref{eq:admquantity} to the boundary flux at future infinity.
A closer look at \eqref{eq:admquantity} leads to a further distinction compared to recent results on evaporating black holes in an AdS setting such as \cite{AlmheiriEMM}. It is in general clear that allowing for non-equilibrium states should correspond to a dynamically evolving ADM quantity. For an asymptotically AdS$_2$ black hole it is reasonable to allow either a time-dependence of the Schwarzian or of the renormalised dilaton value in \eqref{eq:admquantity}. Whereas the former choice corresponds to a time-dependent temperature and hence a dynamically evolving black hole mass as in \cite{AlmheiriEMM}, the latter amounts to a fixed temperature with a dynamically evolving angle (or equivalently a time-dependent dilaton) in three dimensions as in \cite{Verheijden}. The former choice is not an option for a fixed de Sitter background, since a time-dependent temperature does not correspond to a change in integration constant but instead amounts to a shift away from a de Sitter geometry. Hence, we will consider the temperature to be fixed as in \eqref{eq:temperature} and let the dilaton acquire dynamical behaviour. 

\section{Adding matter: a dynamical boundary dilaton}\label{sec:Adding matter:adynamicalboundarydilaton}
In this section we will consider adding matter to our configuration. In this way, we consider the observer inside the static patch to experience an incoming (net) positive energy flux. Solving the equations of motion for the boundary dilaton $\Phi_r \sim \alpha$ at $\mathcal{I}^{+}$, we find a time-dependent $\alpha (u)$. Indeed, from the three-dimensional perspective we see that this corresponds to a shrinking gravitational system in the $\varphi$-direction: the cosmological horizon is evaporating. We also comment on the timescales relevant to our problem.

\subsection{Matter and the Unruh state}\label{sec:Unruhstate}
We wish to consider a setup in which the size of the horizon decreases, such that the entropy \eqref{eq:ourentropy} dynamically evolves. To be able to do so, we will have to specify a quantum state for the radiation. The state we want to consider is the Unruh-de Sitter state established in \cite{backreactionFuture}. Since this state is less discussed in the literature, we will carefully define its construction.

\begin{figure}
\centering
\begin{tikzpicture}
	\pgfmathsetmacro\myunit{4}
	\draw	(0,0)			coordinate (a)
		--++(90:\myunit)	coordinate (b)
							node[pos=.5, above, sloped] {\text{North Pole}}
		--++(0:\myunit)		coordinate (c)
							node[pos=.5, above] {$\mathcal{I}^+$}
		--++(-90:\myunit)	coordinate (d)	
							node[pos=.5, sloped, above] {\text{South Pole}}
		--cycle 			node[pos=.5, below] {$\mathcal{I}^-$};
 	\draw[dashed, name path = hor1] (a) --++ (c);
 	\draw[dashed, name path = hor2] (b) -- (d);
 	\draw[dashed, fill=orange!50!white, fill opacity = 0.5, name intersections={of=hor1 and hor2, by={int1}}] (d) -- (int1) -- (c);
 	\draw[dashed, fill=lightblue, fill opacity = 0.8] (b) -- (int1) -- (c);
 	\draw (a) -- (b);
 	\draw (c) -- (d);
 	\draw (b) -- (c);
 	\draw[orange, decorate, decoration=snake, ->] (3.1,1) --++(45:0.8);
 	\draw[orange, decorate, decoration=snake, ->] (3.4,2.1) --++(135:0.8);
 	\draw[blue, decorate, decoration=snake, ->] (2.5, 2.8) --++ (45:0.8);
 	\draw[blue, decorate, decoration=snake, ->] (1.5, 2.8) --++(135:0.8);
 	\draw[red, decorate, decoration=snake, ->] (2.4,1.2) --++(45:0.8);
 	\draw[red, decorate, decoration=snake, ->] (2.5, 2) --++(135:0.8);
 	\draw[black, thick] (2.8,1.4)--++(135:0.4);
 	\draw[black, thick] (1.1,2.9)--++(45:0.4);
 	\draw[black, thick] (3,2.2)--++(45:0.4);
\end{tikzpicture}
\caption{Penrose diagram of the Unruh state. The black bars denote zero one-point functions, whereas the arrows denote non-zero one-point functions of the stress tensor. Globally (red radiation), there are only left-moving modes and no right-moving modes. In the static patch (orange radiation), this corresponds to no outgoing radiation. This gets transferred to the Milne patch (blue radiation). As also elaborated upon in the main text, in global or Kruskal coordinates there is a flux of negative energy. }
\label{fig:Unruhstate}
\end{figure}
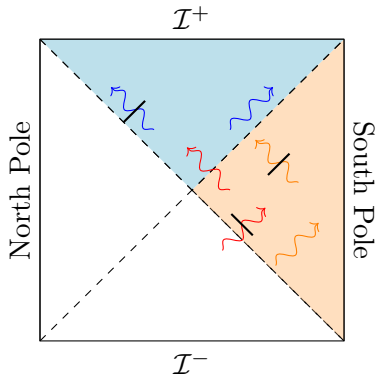

As we are working in a semi-classical limit, we promote the stress tensor components to their expectation values 
$T_{\mu \nu}= \langle T_{\mu \nu} \rangle$. In a curved background, the stress tensor receives contributions from the Weyl anomaly. Our first task is to specify the components of the quantum stress tensor independently of the state. One approach is to demand conservation of the stress energy tensor as in \cite{PhysRevD.15.2088,Lohiya_1978}
\begin{equation}
    \nabla_{\mu} \langle T^{\mu \nu}\rangle=0\,,
\end{equation}
which allows to solve for the components in the general lightcone coordinates of \eqref{eq:generallightconecoordinates}
\begin{align}
    \langle T_{x^\pm x^\pm }(x^{\pm}) \rangle &=\frac{c}{12 \pi}\left(\partial_{\pm}^2 \omega-\partial_{\pm}\omega\partial_{\pm}\omega \right)-\frac{c}{24 \pi} t_{x^\pm x^\pm} (x^{\pm})+\langle \tau_{x^\pm x^\pm }\rangle\label{eq:EMToffdiagonal}\,,\\
    \langle T_{x^+ x^-}(x^{+},x^{-}) \rangle &=-\frac{c}{12 \pi}\partial_{+}\partial_{-}\omega \label{eq:EMTdiagonal}\,.
\end{align}
Here, $\tau_{\mu \nu}$ refers to the (state-independent) contribution to the stress tensor in flat space. While the off-diagonal component (in lightcone coordinates) is completely fixed by the conformal anomaly, the diagonal components include state-dependent contributions: the two independent degrees of freedom $t_{x^\pm x^\pm} (x^{\pm})$. The stress tensor naturally obeys the anomalous transformation law, which in our conventions is\footnote{Note the non-standard minus for the Schwarzian transformation law and non-standard normalisation.}
\begin{equation}\label{eq:schwarzian}
    \langle T_{y^\pm y^\pm }(y^{\pm})\rangle=\left(\frac{d x^{\pm}}{d y^{\pm}}\right)^2 \langle  T_{x^\pm x^\pm }(x^{\pm})\rangle - \frac{c}{24 \pi} \{x^{\pm},y^{\pm} \}\,,
\end{equation}
with the functions $t_{x^\pm x^\pm} (x^{\pm})$ changing accordingly,
\begin{equation}\label{eq:tplusplustminusminus}
 t_{y^\pm y^\pm} (y^{\pm})= \left(\frac{d x^{\pm}}{d y^{\pm}}\right)^2 t_{x^\pm x^\pm} (x^{\pm})-\{x^{\pm},y^{\pm} \}\,.
\end{equation}
Specifying $t_{x^\pm x^\pm} (x^{\pm})$ amounts to fixing a choice of vacuum, and hence a choice of thermal flux for the static patch observer. This determines the flux at future infinity $\mathcal{I}^{+}$:
\begin{equation}\label{eq:flux}
    \langle T_{x^+ x^+ }\rangle-\langle T_{x^- x^- }\rangle=\frac{c}{24 \pi}\left( t_{x^- x^-}-t_{x^+ x^+} \right)\,.
\end{equation}
By fixing the two independent degrees of freedom, we can recover the standard Bunch-Davies vacuum, which is defined with respect to the Kruskal coordinates \eqref{eq:KruskalWatse} to be
\begin{equation}\label{eq:BunchDavies}
     \langle T_{x^\pm x^\pm }(x^{\pm})\rangle = 0\,.
\end{equation}
Note that by use of \eqref{eq:schwarzian} and \eqref{eq:Kruskaltostatic} we can see that the Bunch-Davies state \eqref{eq:BunchDavies} corresponds to thermal equilibrium on the static patch:
\begin{equation}\label{eq:staticequilibrium}
    \langle T_{\sigma^\pm \sigma^\pm}(\sigma^{\pm})\rangle=\frac{\pi c}{12 \beta_{\text{dS}}^2}\,.
\end{equation}
At future infinity \eqref{eq:BunchDavies} corresponds to a zero net flux \eqref{eq:flux}. Hence we must follow the approach of \cite{backreactionFuture} and break the symmetry between incoming and outgoing flux on the static patch or equivalently allow for a net flux \eqref{eq:flux} at $\mathcal{I}^{+}$. Our desired state corresponds to setting the vacuum with respect to the static coordinates for the left-moving and with respect to the Kruskal coordinates for the right-moving modes. Hence in static coordinates we find
\begin{equation}\label{eq:UnruhforgeneralCFTstatic}
    \langle T_{\sigma^+ \sigma^+ }(\sigma^+)\rangle=0\,,\qquad \langle T_{\sigma^- \sigma^- }(\sigma^-) \rangle=\frac{\pi c}{12 \beta_{\text{dS}}^2}\,,
\end{equation}
whereas in Kruskal coordinates this gives
\begin{equation}\label{eq:UnruhforgeneralCFTglobal}
    \langle T_{x^+ x^+ }(x^+)\rangle=-\frac{c}{48 \pi (x^+)^2} \,,\qquad \langle T_{x^- x^- }(x^-) \rangle=0\,.
\end{equation}
Note that the stress tensor \eqref{eq:UnruhforgeneralCFTglobal} is singular at the past horizon and the NEC is violated as required by Hawking's area theorem \cite{Hawkingarealaw,Bousso:2015mqa}. In \figref{fig:Unruhstate} we summarise the fluxes in different patches. 

Since $\mathcal{I}^{+}$ is sensitive to the diagonal components of the stress tensor, it is sensitive to the flux \eqref{eq:flux} and hence also to the state of the quantum fields. By expressing the boundary matter action in terms of $x(u)$ we can see how \eqref{eq:Schwarzianeom} is modified (in addition, an intuitive derivation is given in \cite{MaldacenaNAdS}). For classically conformal matter we arrive at 
\begin{equation}\label{eq:generaldilatonfluxequation}
\frac{1}{8 \pi G_N}\left(\Phi_r \{x(u),u \}'+ 2 \Phi_r' \{x(u),u \}+\Phi_r''' \right)=\langle T_{ x^- x^- }(u) \rangle-\langle T_{ x^+ x^+ }(u) \rangle\,.
\end{equation}
Here, we have a general differential equation relating the change of a previously conserved quantity \eqref{eq:admquantity} to a flux leaving $\mathcal{I}^{+}$, expressed in general lightcone coordinates. Now, as elaborated upon in section \ref{sec:Boundaryaction}, we are specifically interested in the solution \eqref{eq:thermalx} since the corresponding ADM quantity is related to the entropy of the static patch. Hence a flux as in \eqref{eq:generaldilatonfluxequation} will give the desired evolution of the entropy. Therefore, we must actually consider the matter contribution to \eqref{eq:generaldilatonfluxequation} with respect to the Milne coordinates. Due to the analytic continuation linking Milne and static coordinates, the stress tensor in terms of Milne coordinates is also given by \eqref{eq:UnruhforgeneralCFTstatic}.
Note that in \eqref{eq:generaldilatonfluxequation} the stress tensor components are given in terms of the boundary variable $u$. 
Therefore, if we now use \eqref{eq:schwarzian}, assume both the state \eqref{eq:UnruhforgeneralCFTglobal} and no initial outgoing matter, for $u>0$ we deduce \cite{HermanBackreaction}
\begin{align}\label{eq:ourboundarystate}
  \langle  T_{ x^+ x^+ }(u) \rangle &=0\,, \\ 
  \langle  T_{ x^- x^- }(u) \rangle &=-\frac{c}{24 \pi}\{x,u \}\,,
  \end{align}
and therefore also
\begin{equation}
\frac{1}{8 \pi G_N}\left(\Phi_r \{x(u),u \}'+ 2 \Phi_r' \{x(u),u \}+\Phi_r''' \right)=-\frac{c}{24 \pi}\{x(u),u \}\,.
\end{equation}
For de Sitter spacetimes the temperature is an intrinsically fixed quantity, such that we may work with the simpler equation
\begin{equation}\label{eq:dilatoneom}
\left(-4 \frac{\pi^2}{\beta_{\text{dS}}^2}  \Phi_r' + \Phi_r''' \right)=\frac{2\, c\, G \pi^2 }{3 \, \beta_{\text{dS}}^2}\,.
\end{equation}
In principle \eqref{eq:dilatoneom} yields both homogeneous exponential and linear inhomogeneous solutions. As we are interested in the backreaction of the matter on the dilaton we work with the latter. Hence, we solve \eqref{eq:dilatoneom} as 
\begin{equation}\label{eq:dilatonsolution}
    \Phi_r (u) =2 \pi \ell\left(1-\frac{c G}{12 \pi \ell}u\right)\,.
\end{equation}
Here, we have imposed the condition $\Phi_r (u = 0) = 2 \pi \ell$. We have determined the backreacted, renormalised dilaton value in terms of the Euclidean boundary time $u$ of the quantum mechanical model living at future infinity.\footnote{In the language of \cite{AlmheiriOutside}, the degree of freedom of the dot.} This means that we are reducing the dynamics of the gravity+CFT system living on two-dimensional de Sitter to the dynamical boundary function \eqref{eq:dilatonsolution}. Let us now make the connection to the higher-dimensional picture of \figref{fig:Setup}. Note that \eqref{eq:dilatonsolution} corresponds to
\begin{align}\label{eq:anglesolution}
    \alpha(u)=\left(1-\frac{c G}{12 \pi \ell}u\right)\,.
\end{align}
Hence, at least at $\mathcal{I}^{+}$ we see that \eqref{eq:dilatonsolution} may be understood as transparent boundary conditions for the flux moving along $\varphi$.
Different phases of evaporation correspond to the evolution of the parameter $\alpha$ as given in \eqref{eq:anglesolution}. From \eqref{eq:anglesolution} we determine the Page time, i.e., the value of $u$ for which $\alpha$ equals $1/2$:
\begin{equation}\label{eq:pagetime}
    u_{\text{Page}}=\frac{6 \pi \ell}{c G}\,.
\end{equation}
Moreover, \eqref{eq:anglesolution} may also be understood as the evolution of the inverse of an effective Hubble parameter $\ell \dot\alpha(u)$,
 \begin{align}\label{eq:effectivehubbleparameter}
   \ell \dot\alpha(u)=-\frac{c}{6}\frac{1}{S_{\text{dS}}}\,,
\end{align}
such that the backreaction is suppressed by the entropy. 

To conclude, we see that even though we set evaporating conditions on the two-dimensional spacetime, due to the nature of the dilaton, determining  the backreaction on $\Phi_r$ immediately implies dynamical evolution in the three-dimensional description.
    
\subsection{Estimates on the de Sitter lifetime}\label{sec:FirstEstimatesondeSitterLifetime}
Any approach to de Sitter space should take into account the restrictions imposed on specific observers. More concretely, as we are interested in determining the evolution of the entropy of the radiation, we should always compare with possible bounds on the lifetime of de Sitter space as these might constrain up to which point we can trust the entropy computations of section \ref{sec:entropies}. Here we give a general argument before moving to a bulk description in section \ref{sec:BackreactionConsiderationsandTimescales}. 

The least restrictive timescale is the recurrence or Poincar\'e time. Following \cite{troublewithds}, we can view de Sitter in thermal equilibrium as a thermofield double state in analogy with the ideas of \cite{Maldacena:2001kr}. However, it can be shown that the assumption of finite entropy contradicts the algebra acting on the thermofield double state,\footnote{It may be also argued that for this reason for a single observer the Hilbert space only describes one side of the horizon. Only the horizon-invariant subalgebra would correspond to physical states \cite{Parikh:2004wh}.} which implies that the symmetry between different static patches is broken. This introduces a new timescale
\begin{equation}\label{eq:recurrencetime}
    t \sim \exp(S_{\text{dS}}) ~,
\end{equation}
indicating when the space may not be approximated by de Sitter anymore. However, as elucidated in the previous section, we are using an out-of-equilibrium state in which the de Sitter isometries are broken from the onset. This should drastically reduce any lifetime considerations to a timescale smaller than \eqref{eq:recurrencetime}. It would be interesting to consider in detail how the argument leading to \eqref{eq:recurrencetime} has to be modified.

In \cite{backreactionFuture} a bound on the lifetime in the Unruh-de Sitter state was given as
\begin{equation}
    t  \sim S_{\text{dS}}\,.
\end{equation}
 As also stated above, in this low dimensional setting we may think of \eqref{eq:anglesolution} as determining the evolution of an effective Hubble parameter \eqref{eq:effectivehubbleparameter}. As we recover the same behaviour demonstrated in \cite{backreactionFuture} for the effective Hubble parameter, we consider the same bound. Hence, in our language the lifetime on dS in the Unruh state is given as
\begin{equation}\label{eq:Unruhlifetime}
    u \sim u_{\text{Page}}\,,
\end{equation}
where we specified $u_{\text{Page}}$ in \eqref{eq:pagetime}. From the boundary perspective it might not be quite clear what effect should actually constrain the system to this timescale. However, in section \ref{sec:BackreactionConsiderationsandTimescales} we will use a specific bulk argument first used in \cite{Aalsma:2021bit}, which demonstrates the appearance of a trapped region at the time \eqref{eq:Unruhlifetime}.

\section{The static patch}\label{sec:staticpatch}
So far we have restricted ourselves to the use of boundary calculations. However, as the static patch is `disconnected' from future infinity by a cosmological horizon, it might not be immediately clear to what extent \eqref{eq:anglesolution} may be applied inside the static patch and where exactly a thermal bath should be located for the static observer. We should therefore understand the properties of the dilaton inside the static patch. We will locate a non-gravitating region for which the dilaton notably exhibits the same behaviour as at $\mathcal{I}^{+}$, which will justify the calculations of section \ref{sec:staticentropy}.

\subsection{The bulk dilaton solution}
Let us start by stating the backreacted bulk dilaton solution. In JT gravity, backreaction effects are fully captured by the dilaton, such that we have to solve the equations \eqref{eq:BulkEoms} with the sources \eqref{eq:EMTdiagonal} and \eqref{eq:UnruhforgeneralCFTglobal}.
Solving the set of differential equations gives
   \begin{equation}\label{eq:backreacteddilaton}
   \begin{aligned}
       \Phi (x^+,x^-) &= \frac{a_1 x^- + a_2 x^+}{\ell^2 - x^+ x^-} + a_3 \left(1 - \frac{2\ell^2}{\ell^2 - x^+x^-} \right) \\
       & +\frac{c G}{6} \left( \frac{2\ell^2}{\ell^2 - x^+x^-} - \frac{\ell^2 +x^+ x^-}{\ell^2-x^+x^-} \log{\frac{x^+}{\ell}} \right)~,
      \end{aligned}
    \end{equation}
where $a_1,a_2,a_3$ are integration constants. We wish to construct a solution that qualitatively matches the structure of \eqref{eq:dilatonsolution}. This means we want to recover the background solution \eqref{eq:KruskalWatse}, with a quantum correction enforcing a decreasing Gibbons-Hawking entropy. Hence, we fix the integration constants to $a_1 = 0 = a_2$ and $a_3 = - 2\pi \alpha + \frac{cG}{6}$. Then we find  
\begin{equation} \label{eq:backreactedbulkdilaton}
    \begin{aligned}
        \Phi (x^+,x^-) = \frac{\ell^2 + x^+x^-}{\ell^2 - x^+x^-} 2 \pi\left( \alpha - \frac{cG}{12\pi} \log\frac{x^+}{\ell} \right) + \frac{cG}{6}~.
      \end{aligned}  
    \end{equation}
The parameter $\alpha$ of the background solution sets the value of the Gibbons-Hawking entropy. The structure in the brackets of \eqref{eq:backreactedbulkdilaton} may be understood as this parameter $\alpha$ acquiring dynamical behaviour due to the backreaction of the quantum state. We can use the rescaling symmetry in $(x^+,x^-)$ to set $\alpha=1$ in \eqref{eq:backreactedbulkdilaton}; then, taking it to $\mathcal{I}^{+}$ we recover the behaviour \eqref{eq:dilatonsolution} and hence also \eqref{eq:anglesolution}. 
The last constant term of \eqref{eq:backreactedbulkdilaton} corresponds to a shift in the vacuum which already occurs for the Bunch-Davies state \cite{Sybesma}. As this merely amounts to a rescaling of $S^1$ it is not interesting for our purposes, and we drop this term.

\subsection{The cosmological horizon and quantum mechanics}
In the static patch, the Unruh state \eqref{eq:UnruhforgeneralCFTstatic} corresponds to incoming radiation. Therefore, a natural set of coordinates to describe the bulk dilaton solution inside the static patch is given by the incoming Eddington-Finkelstein coordinates $(\sigma^+, r)$ with $\sigma^+ = t + r^*$ as in \eqref{eq:staticpatchnullcoordinates}. In these coordinates, the two-dimensional metric is 
    \begin{equation}
        ds^2 = - \left( 1 - \frac{r^2}{\ell^2} \right) (\dd \sigma^+)^2 + 2\dd \sigma^+ \dd r ~.
    \end{equation}
The backreacted bulk dilaton \eqref{eq:backreactedbulkdilaton} with $\alpha = 1$ takes on the form 
    \begin{equation} \label{eq:staticbulkdilaton}
        \Phi (\sigma^+, r) = \frac{r}{\ell} 2\pi \left( 1 - \frac{c G}{12\pi \ell} \sigma^+ \right)~.
    \end{equation}
As can be seen by the appearance of $\Phi$ in \eqref{eq:JTaction2}, in two dimensions $\Phi$ plays the role of the inverse of the gravitational coupling. Hence we can see by the structure of \eqref{eq:staticbulkdilaton} that at the pole $r \to 0$ we arrive at a strongly coupled gravitational region. On the other hand, at the past cosmological horizon $\sigma^+ \to -\infty$, we find that gravity becomes weak. Hence, while naturally $\mathcal{I}^{+}$ plays a special role as gravity fully decouples there, here we have located a second region inside the bulk for which the same logic holds. This weakly-gravitating or non-dynamical region will allow us to compute the entropy of radiation collected inside the static patch \cite{Raju:2021lwh}. 

From \eqref{eq:staticbulkdilaton} we can read off the backreacted value of $\alpha$ in the static patch (denoted $\alpha_s$) by comparing to \eqref{eq:bulkdilatondimred} (with $\phi = \frac{r}{\ell}$) 
    \begin{equation} \label{eq:bulkalpha2}
        \alpha_{\rm s} (\sigma^+) = 1 - \frac{c G}{12 \pi \ell} \sigma^+~.
    \end{equation}
We see that we recover exactly the behaviour \eqref{eq:dilatonsolution}, but with the spacelike parameter $u$ replaced by the null coordinate $\sigma^+$. For an observer at the pole $(r = 0)$, this reduces to genuine time-dependence as $\sigma^+ (r = 0) = t$: 
    \begin{equation} \label{eq:bulkalphastatic}
        \alpha_{\rm{s, pole}} (t) = 1 - \frac{c G}{12\pi \ell} t~.
    \end{equation}
Note that while it is clear that merely taking the bulk solution \eqref{eq:backreactedbulkdilaton} to the boundary $\mathcal{I}^+$ must give the same behaviour, as guaranteed by the equivalence of the Schwarzian description to the bulk equations \eqref{eq:BulkEoms}, the behaviour exhibited in \eqref{eq:bulkalpha2} and \eqref{eq:bulkalphastatic} is more pronounced. We can view \eqref{eq:bulkalphastatic} as the static patch counterpart of \eqref{eq:dilatonsolution}, and it is then tempting to think of this as the timelike realisation of the quantum mechanical model living at future infinity. This might be in line with a `stretched horizon' static patch holography \cite{Susskind:2021omt,SusskindStretched}, where the physics inside the static patch (i.e., the physics as experienced by an observer at $r=0$) has a dual description in which the degrees of freedom are located at the `boundary' of the static patch, i.e., the horizon. As the evaporation angle $2\pi \alpha \in [0,2\pi)$ we see that the evaporation process is finite: the observer at the pole can collect radiation only for a finite time. 

This is indeed the interpretation we will take: whereas it is the static patch observer at $r=0$ who is collecting radiation, thereby invoking the dynamical behaviour of $\alpha$, there is a dual description at the weakly-coupled past horizon. This, then, is the region where we will compute the entanglement entropy of the emitted radiation via the auxiliary, three-dimensional spacetime, where a dynamical $\alpha$ corresponds to a dynamical interface in the $\varphi$-direction. We will recap the logic in the next section, where we perform the aforementioned calculations. 



\section{Fine-grained entropy calculations} \label{sec:entropies}
We are now in a position to calculate the fine-grained entropy of the radiation as a function of $u$ at $\mathcal{I}^{+}$ and as a function of $t$ at the past cosmological horizon in the static patch. These are the two regions of the two-dimensional de Sitter spacetime at which the gravitational dynamics reduce to quantum mechanical descriptions. At the two aforementioned decoupling regions we encounter a description in terms of a matter CFT coupled to the dilaton determined by backreaction. In our construction, we consider the dilaton to have a higher-dimensional origin, such that the transparency relates to the spherical coordinate $\varphi$; this functions as an auxiliary system. This allows the calculation of the fine-grained entropy of the two-dimensional system at the gravitationally decoupled regions to be performed via this auxiliary system. For both regions we use the higher-dimensional setting to first calculate the entanglement entropy in thermal equilibrium of a subregion with interval $\Delta \varphi$ and then by use of the solutions \eqref{eq:anglesolution} and \eqref{eq:bulkalpha2} we imbue the results with dynamical evolution determined by the backreaction, to obtain the desired out-of-equilibrium state.
       
\subsection{Entropy computed at \texorpdfstring{$\mathcal{I}^+$}{I+}} \label{sec:futureentropy}
\begin{figure}
    \centering
\begin{tikzpicture}
    \draw[very thick, green!40!black] (0,0) arc[start angle=0, end angle=260, radius = 2cm] coordinate (A);
    \filldraw[red!80!black] (0,0) circle(0.03cm);
    \filldraw[red!80!black] (A) circle (0.03cm);
    \draw[very thick, red!80!black, opacity=0.5] (0,0) arc[start angle=0, end angle=-100, radius=2cm] ;
    \draw[<->, gray] (-0.2,0) arc[start angle=0, end angle=-100, radius=1.8cm] node[pos=.6, above=.2cm] {\small{$\alpha (u)$}};
    \path (A) --++(0:0.05) coordinate (R1);

    \draw[blue, decorate, decoration={snake, amplitude=0.6mm, segment length=0.3cm}, ->] (A) arc[start angle=260, end angle=235, radius = 2cm] coordinate (R2);
    \path (R2) arc[start angle=235, end angle = 215, radius = 2cm] coordinate (R3);
    \draw[blue, decorate, decoration={snake, amplitude=0.6mm, segment length=0.3cm}, ->] (R3) arc[start angle=215, end angle=190, radius = 2cm] coordinate (R4);
    \draw[blue, decorate, decoration={snake, amplitude=0.6mm, segment length=0.3cm}, ->] (0,0) arc[start angle=0, end angle=25, radius= 2cm] coordinate (R5); 
    \path (R5) arc[start angle=25, end angle = 50, radius = 2cm] coordinate (R6);
    \draw[blue, decorate, decoration={snake, amplitude=0.6mm, segment length=0.3cm}, ->] (R6) arc[start angle=50, end angle=75, radius= 2cm] coordinate (R7);
\end{tikzpicture}
    \caption{A constant $\chi$ slice of the (renormalised) cylinder at $\mathcal{I}^{+}$ of three-dimensional de Sitter in our partial reduction approach. In the two-dimensional picture the radiation ends up at $\mathcal{I}^{+}$ (red), before evaporating into the bath (green), which is located along a higher dimension ($\varphi$). The transfer of radiation from dynamical gravity to bath corresponds to \eqref{eq:anglesolution}. Note that as $u$ increases, $\alpha (u)$ will decrease such that we are indeed modelling an evaporating system. We included the higher-dimensional de Sitter region in opaque red to clarify the entanglement structure.}
    \label{fig:futureinfinityevaporationpicture}
\end{figure}
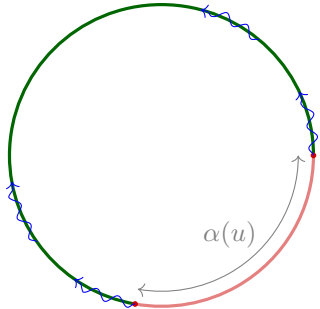

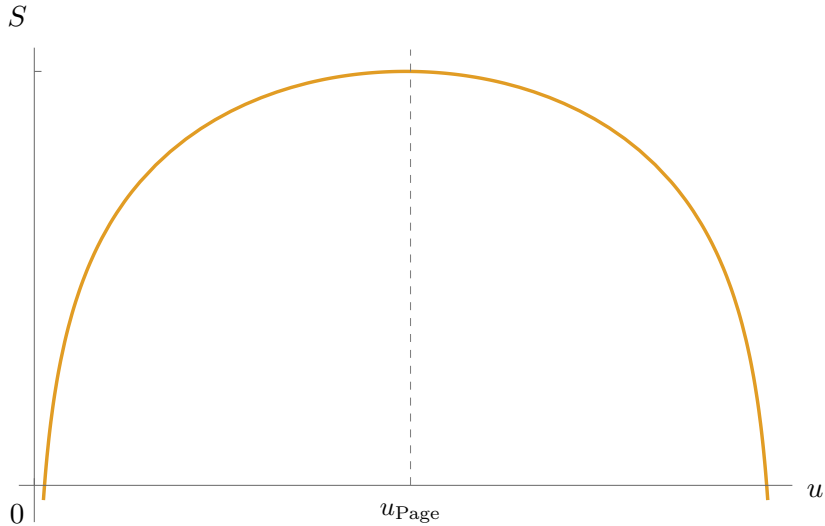
\begin{figure}
        \centering
    \begin{tikzpicture}[scale=1.5]
	\node[anchor=south west, inner sep=0] at (0,0) {\includegraphics[width=0.675\textwidth]{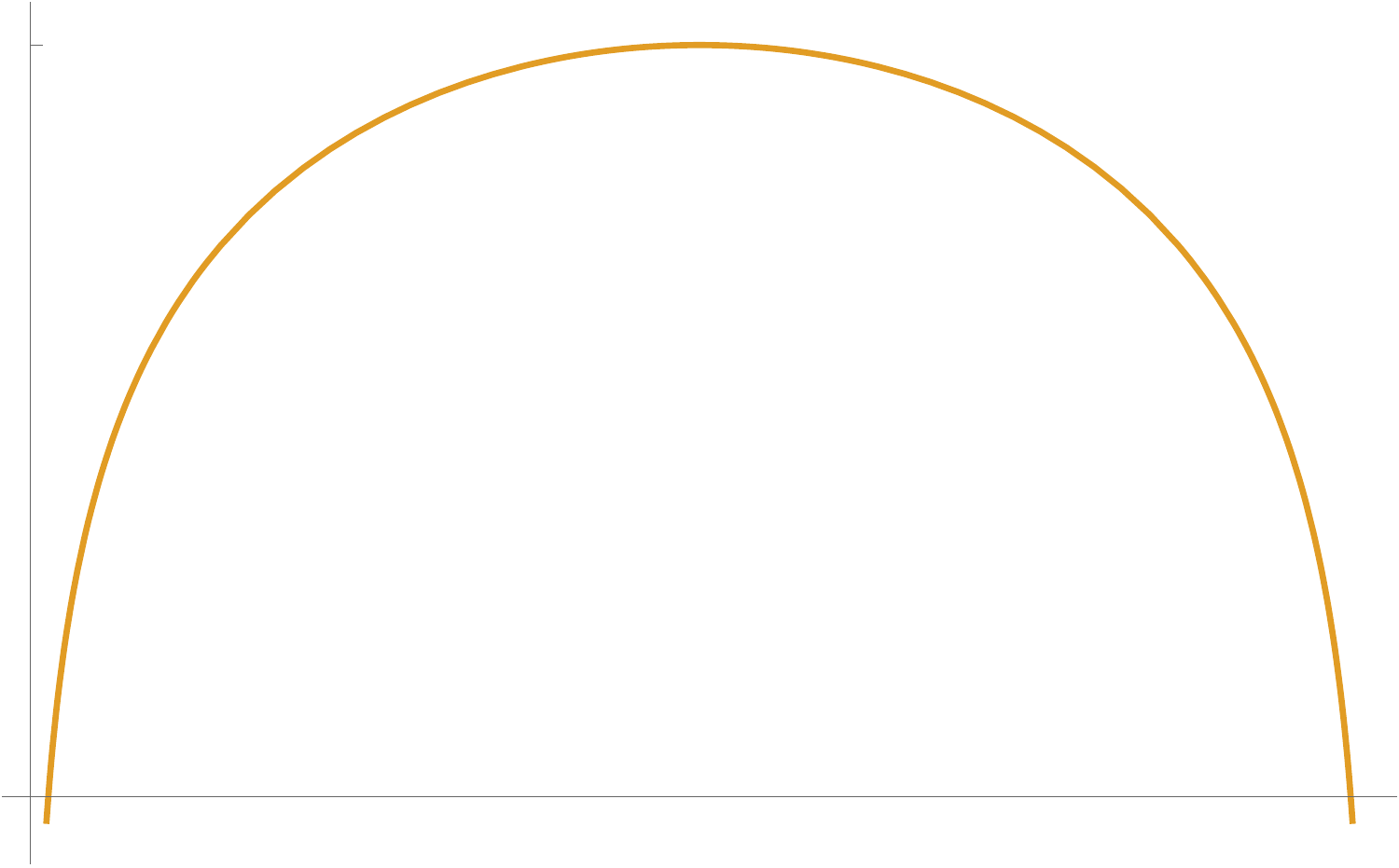}};
	\node at (0,0.1) {$0$};
	\node at (-0.2,4.01) {};
	\node at (7,0.3) {$u$};
	\node at (0,4.5) {$S$};

    \draw[gray, dashed] (3.45,0.35)--(3.45,4.2);
	\node at (3.45,0.1) {\small{$u_\text{Page}$}};
    \end{tikzpicture}
        \caption{The radiation entropy collected at $\mathcal{I}^{+}$. For this plot, we took the cut-off $\epsilon = 2 \ell e^{-2 G S_\text{dS}}$, and $\ell = G = 1$, which fixes $c = \frac{3}{2}$. The qualitative behaviour is the same for any $c$; this just determines the range of $u$.}
        \label{fig:futureinfinity}
    \end{figure}

Consider the three-dimensional Milne solution
\begin{equation}\label{eq:3dMilne}
    ds^2 = - \dd \tau^2 + \sinh^2\frac{\tau}{\ell} \dd \chi^2 + \ell^2 \cosh^2\frac{\tau}{\ell} \dd \varphi^2~,
\end{equation}
and its partial reduction \eqref{eq:MilneDilaton}. The gravitational dynamics in the reduced de Sitter space \eqref{eq:MilneDilaton} with the coordinates $(\tau,\chi)$ reduce to that of a quantum mechanical boundary degree of freedom, namely \eqref{eq:anglesolution}. Note that this is in accordance with a dS/CFT picture \cite{Stromingerdscft}. At future infinity the induced metric is
 \begin{equation}\label{eq:Milneatfutureinfinity}
            ds^2 = \frac{1}{4} e^{2\tau^*/\ell} \lp \dd \chi^2 + \ell^2 \dd \varphi^2 \rp~,
    \end{equation}
where $\tau^*$ denotes the asymptotic value $\tau \to \tau^* = \infty$. The setup at $\mathcal{I}^{+}$ is depicted in \figref{fig:futureinfinityevaporationpicture}.
As we want to use the general formula for entanglement entropy on curved spacetimes Weyl-equivalent to flat spacetime, we define new coordinates
\begin{equation}\label{eq:xxbardefinitions}
        x = e^{\xi + i \varphi}~, \quad \bar{x} = e^{\xi - i \varphi}~, 
    \end{equation}
where $\xi= \chi/\ell$. In these coordinates, \eqref{eq:Milneatfutureinfinity} is given by
    \begin{equation}
        ds^2 = \frac{\ell^2}{\Omega^2} \dd x \dd\bar{x}~, \qquad \Omega = 2 e^{-\tau^*/\ell} \sqrt{x \bar{x}}~.
    \end{equation}
The matter (CFT) entropy in a region with general endpoints $(x_1, \bar{x}_1)$ and $(x_2, \bar{x}_2)$ now is 
    \begin{equation} \label{eq:3Dentropy}
    \begin{aligned} 
        S_\text{mat} &= \frac{c}{6}\log \left[ \frac{(x_1 - x_2)(\bar{x}_1-\bar{x}_2)}{ \Omega (x_1) \Omega (x_2)} \right]\\
        &=\frac{c}{6}\log \left[\frac{e^{2  \tau^*/\ell}}{4}\left(2 \cosh{(\xi_1-\xi_2)}-2\cos(\varphi_1-\varphi_2) \right)   \right]\,.
    \end{aligned}
    \end{equation}
We are interested in calculating the matter entropy in the Unruh state \eqref{eq:UnruhforgeneralCFTglobal}, but from the three-dimensional point of view we can apply the standard formula in thermal equilibrium as in \eqref{eq:3Dentropy}. The net flux corresponding to the Unruh state in 2D arises by imposing $\alpha (u)$ as in \eqref{eq:anglesolution}. Since in the auxiliary system we only consider separation in the direction of the dimensional reduction, we take  $\chi_1 = \chi_2$ and $\Delta \varphi = 2\pi  (1-\alpha)$, such that we get
     \begin{equation}
    \begin{aligned}\label{eq:Evitafutureinfinityformula1}
        S_\text{rad} &= \frac{c}{6}\log \left[ e^{2\tau^*/\ell} \sin^2 \Delta \varphi \right] \\
        &= \frac{c}{6}\log \left[ \frac{4\ell^2 }{\epsilon^2} \sin^2\frac{\Phi_r}{2 \ell} \right]\,,
     \end{aligned}
    \end{equation}
where we collected the coordinate-dependent UV-divergences as $\epsilon = 2\ell e^{-\tau^*/\ell}$, in line with $\Phi_b = \frac{\Phi_r}{\epsilon}$ for the Milne coordinates \eqref{eq:MilneDilaton}. Writing \eqref{eq:Evitafutureinfinityformula1} as a function of $u$, we find 
    \begin{align}\label{eq:Evitafutureinfinityformula2}
       S_\text{rad}&=\frac{1}{2 G}\log \sin \frac{\Phi_r (u)}{2\ell} + \frac{1}{2G}\log \frac{2\ell}{\epsilon} \nonumber\\
       &=\frac{1}{2 G}\log\sin \pi \left(1-\frac{cG}{12 \pi \ell }u\right) +\frac{1}{2G}\log \frac{2\ell}{\epsilon}\,,
    \end{align}   
where we used $\frac{c}{3} = \frac{1}{2 G^{(2)}}$. Plotting \eqref{eq:Evitafutureinfinityformula2} with an appropriate cut-off gives \figref{fig:futureinfinity}.

Let us postpone backreaction considerations and only comment on \figref{fig:futureinfinity} for the moment. We can see that the entropy reaches the highest point at the Page time \eqref{eq:pagetime}, after which it decreases. We take this to mean that the meta-observer located at $\mathcal{I}^{+}$ observes the evaporating geometry as a pure state. Moreover, note that we did not apply the formula \eqref{eq:island} or demanded purity of some specific system. We arrive at a pure state of radiation without use of any involved argument. As a meta-observer at $\mathcal{I}^{+}$ is located behind the cosmological horizon, and thus has access to the entire history of their universe, for such an observer there is no naive division between interior and exterior subsystems. Hence, the formula \eqref{eq:island} does not have to be applied. 
 
However, we have pointed out a restriction due to the finite lifetime of this state in section \ref{sec:FirstEstimatesondeSitterLifetime}. We will show in section \ref{sec:BackreactionConsiderationsandTimescales} that a trapped surface forms at the Page time. As this complies with geodesic incompleteness, we do not consider the entropy curve after the Page time to be observationally meaningful as the radiation would not reach $\mathcal{I}^{+}$. More accurately, it is most likely not even appropriate to speak of $\mathcal{I}^{+}$ after $u_{\text{Page}}$ anymore due to the occurrence of a singularity. Hence, the meta-observer will not recover information. We still find the comments above on the role of the meta-observer valuable in the larger context of different observers in de Sitter spacetimes and the use of the island formula.

\subsection{Entropy computed inside the static patch} \label{sec:staticentropy}
Next, we will consider an observer at $r=0$ collecting radiation coming from the past cosmological horizon; this corresponds indeed to the Unruh state, and evokes a dynamical backreacted dilaton \eqref{eq:staticbulkdilaton}. As explained in section \ref{sec:staticpatch}, the two-dimensional gravitational system coupled to conformal matter on the reduced metric \eqref{eq:staticpatchnullcoordinates} can be described by a single degree of freedom close to the past cosmological horizon: the renormalised dilaton $\Phi_r (\sigma^+) = 2\pi \alpha (\sigma^+)$, where $\alpha (\sigma^+)$ is given in \eqref{eq:bulkalpha2}. In the three-dimensional geometry \eqref{eq:3dstaticpatch}, this describes transparency in the auxiliary $\varphi$-direction. Thus, we will consider the entanglement entropy of a subsystem with interval $\Delta \varphi$ in thermal equilibrium before using \eqref{eq:bulkalphastatic} to imbue the time-dependence as seen by a static observer. 

%

Recall that the collection of radiation happens at $r=0$, i.e., in a thermal state, but we will take a stretched horizon point of view in which we use the `dual' degree of freedom at the past horizon, which is a gravitationally weakly-coupled region. Within the three-dimensional auxiliary system there is only separation in the $\varphi$-direction between the subsystems, and the induced metric at the past horizon is flat. Thus we may use the standard thermal CFT result \cite{Calabrese:2004eu}: 
\begin{equation}\label{eq:equilibriumentropy}
    S=\frac{c}{3}\log \sinh{\frac{\pi}{\beta_{\text{dS}}}\ell \Delta \varphi} + \frac{c}{3} \log \frac{r }{\ell}\,,
\end{equation}
where $\Delta\varphi$ is a general angle separation and $\beta_{\text{dS}}$ is the inverse of the dS intrinsic constant temperature given in \eqref{eq:temperature}. Note that $\beta_{\text{dS}} = 2\pi \ell$ such that the expression above actually simplifies; here we left $\beta_{\text{dS}}$ to indicate thermal behaviour. The entropy \eqref{eq:equilibriumentropy} may be most conveniently derived by a holographic approach. As such the second term of \eqref{eq:equilibriumentropy} is the UV cut-off defined in the usual holographic manner ($r \to \frac{\ell^2}{\epsilon}$ with $\epsilon \to 0$).

Returning to the two-dimensional perspective, the angular interval of the bath is $\Delta \varphi = 2\pi (1-\alpha)$, and the Unruh state gives a dynamical $\alpha$. 
For the observer at the pole, we will use \eqref{eq:bulkalphastatic}, such that the entropy \eqref{eq:equilibriumentropy} gives (using $\frac{c}{3} = \frac{1}{2 G^{(2)}}$)
    \begin{equation}\label{eq:staticpatchentropy}
    \begin{aligned}
        S_\text{rad} &= \frac{1}{2G} \log \sinh \frac{\pi}{\beta_{\text{dS}}}\frac{c G}{6} t + \frac{1}{2G} \log \frac{\ell }{\epsilon}\\
        &=\frac{1}{2G} \log \sinh  \lp \frac{c G}{ 12 \ell } t \rp + \frac{1}{2G} \log \frac{\ell }{\epsilon}~.
    \end{aligned}
    \end{equation} 
 The expression \eqref{eq:staticpatchentropy} holds before the Page time.
Comparing the expression in the second line of \eqref{eq:staticpatchentropy} to \eqref{eq:Evitafutureinfinityformula2} clearly shows the thermal behaviour of the static patch compared to the non-thermal behaviour at future infinity. We can also see that at later times (before the Page time) the expression \eqref{eq:staticpatchentropy} exhibits linear behaviour in $t$.

    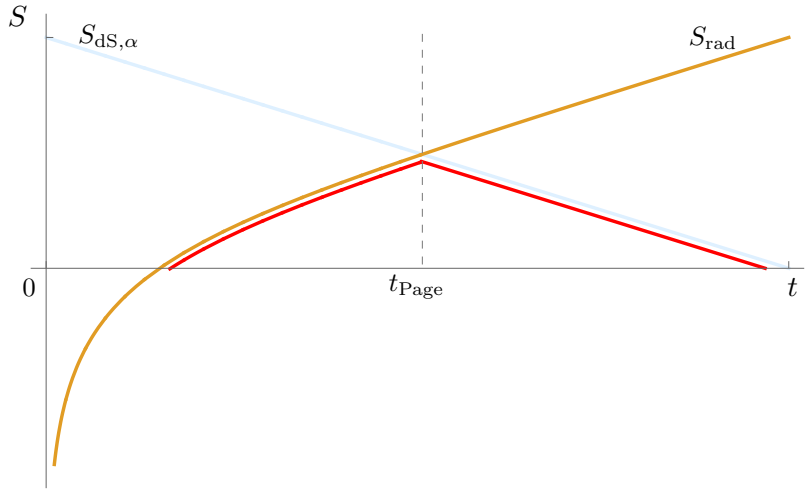
\begin{figure}
        \centering
    \begin{tikzpicture}[scale=1.5]
	\node[anchor=south west, inner sep=0] at (0,0) {\includegraphics[width=0.675\textwidth]{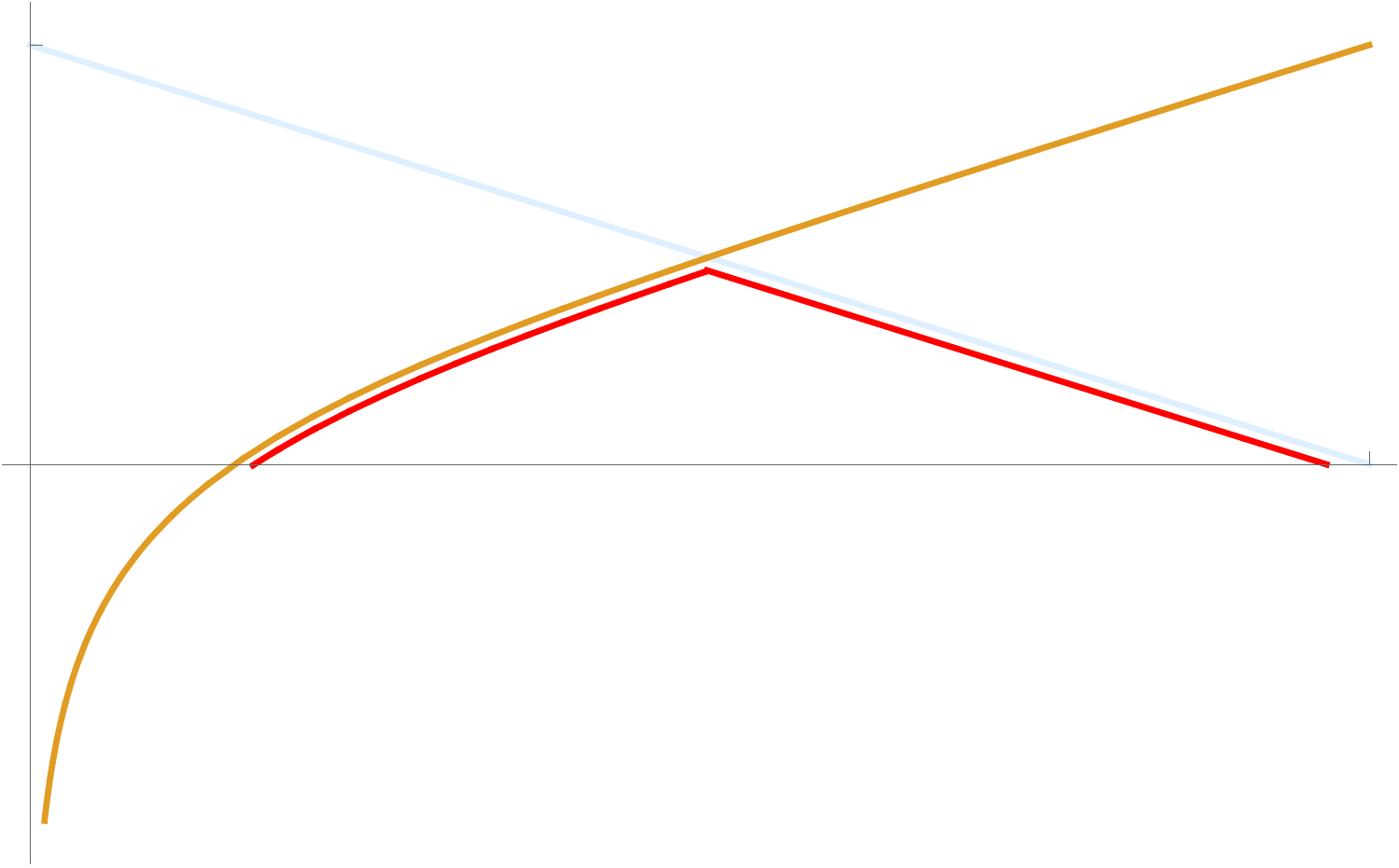}};
	\node at (0,1.8) {\small{$0$}};
	\node at (3.4,1.8) {\small{$t_\text{Page}$}};
	\node at (-0.1,4.2) {$S$};
	\node at (0.7,4) {\small{$S_{\text{dS},\alpha}$}};
	\node at (6,4) {\small{$S_\text{rad}$}};
	\node at (6.7,1.8) {$t$};
	\draw[dashed, gray] (3.45,2.0)--(3.45,4.1);

\end{tikzpicture}
        \caption{The decreasing de Sitter entropy (light blue) and the radiation entropy (orange) as collected inside the static patch. We took the cut-off $\epsilon$ such that the complement of the radiation entropy overlaps at early times with the decreasing de Sitter curve. To plot we set $ \ell = G = 1$; then, $\epsilon = \sinh(\pi) \ell e^{-\pi}$. Note that the negative entropy values at early times are an effect of the finite cut-off and indicate that the computation should not be trusted before this timescale.}
        \label{fig:staticpatchentropy}
    \end{figure}
 While any considerations after the Page time are from a strictly observational perspective irrelevant just as for the meta-observer, they are still useful for our understanding.  As stated previously, we are essentially working on a flat submanifold, with a non-gravitational system coupled to \eqref{eq:bulkalphastatic}, such that we can use holographic arguments for the calculation of this specific entropy. The arguments of \cite{OoguriBao,Verheijden} should also apply for this case; we briefly outline them here. In the large $N$ limit, the entanglement entropy for a two-dimensional CFT at finite temperature can be evaluated using the RT formula in a BTZ geometry. In such a geometry, in general two potential minimal surfaces may be considered candidates for the RT surface as a function of the interval size, which in our language translates to a dependence on $t$. At early times a single connected component contributes, leading to the thermal expression \eqref{eq:staticpatchentropy}. At late times (i.e., after the Page time) a phase transition occurs and there are in principle two disconnected contributions, one of which is disregarded by the demand of purity as in \cite{Verheijden}. Thus, after the Page time we are still working with \eqref{eq:equilibriumentropy} but with the complementary interval 
    \begin{equation}
        \Delta \varphi = 2\pi \alpha (t)\,.
    \end{equation} 
This leads to a unitary Page curve, plotted in \figref{fig:staticpatchentropy}: the static patch observer too would in principle see a pure state. For the static patch observer, this conclusion however requires formula \eqref{eq:island} just as in \cite{Aalsma:2021bit}. Again, in practice catastrophic backreaction forbids information recovery and we should consider the curve to end at the Page time.

\subsection{Backreaction considerations: formation of a trapped region} \label{sec:BackreactionConsiderationsandTimescales}
\begin{figure}
    \centering
    \begin{tikzpicture}[scale=1.3]
	\draw	(0,4)			coordinate (a)
		--++(0:4)	        coordinate (b) node[pos=.5, below] {$u = 0$}
							node[pos=.75, above] {$u_\text{trap}$}
		--++(-90:4)   		coordinate (c) node[pos=.5, right] {$t=0$}
							node[pos=.25, right] {$t_{\text{trap}}$};
	\draw (3.95,2) -- (4.05,2);
	\draw (2,3.95) -- (2,4.05);							
	\draw (3.95,3) -- (4.05,3);
	\draw (3,3.95) -- (3,4.05);
 	\draw[red, sloped, name path = hor2] (c) -- (a) node[pos=.5, below] {$\Phi = \infty$};
 	\draw[dashed] (2,2)--(4,4);
 	\filldraw[thick, draw=lightblue, fill=gray, fill opacity=0.2] (3,4) -- (4,3) node[pos=0.4, left] {$\gamma_1$} to[bend left=35] (4,4) node[pos=.5, left=-1.25] {$\gamma_2$};
 	\draw[draw opacity =0.3] (c) -- (0,0) -- (a);
 	\draw[dashed, draw opacity=0.3] (0,0)--(2,2);
 	\draw (a) -- (4,8) -- (b);
    \draw[orange, decorate, decoration={snake, segment length=0.2cm, amplitude=0.4mm}, ->] (2.3, 1.8) -- (2.7, 2.2);
    \draw[orange, decorate, decoration={snake, segment length=0.2cm, amplitude=0.4mm}, ->] (2.6,1.5) -- (3.0,1.9);
    \draw[orange, decorate, decoration={snake, segment length=0.2cm, amplitude=0.4mm}, ->] (2.5, 4.2) -- (2.9, 4.6);
    \draw[orange, decorate, decoration={snake, segment length=0.2cm, amplitude=0.4mm}, ->] (3.5, 5.2) -- (3.9, 5.6);
    \draw[dashed, lightblue] (3,4) -- (4,5);
    \end{tikzpicture}
    \caption{The lower half of the Penrose diagram is de Sitter space; indicated in red is the past singularity. The upper triangle is Minkowski space, which we drew to indicate the effect of the trapped region (shaded in gray) on what the meta-observer can see. The observer inside the static patch can collect radiation up to the Page time $t_\text{Page} = t_\text{trap}$, at which time the trapped region forms. Similarly, the meta-observer can see radiation from before $u_\text{trapped}$ only.}
    \label{fig:trappedregion}
\end{figure}
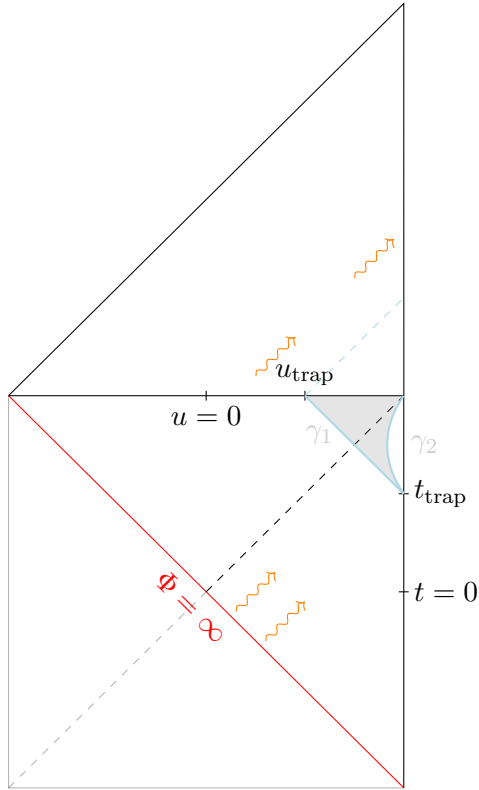

In section \ref{sec:FirstEstimatesondeSitterLifetime} we determined the time of destabilisation for the Unruh-de Sitter state to be set by the Page time \eqref{eq:pagetime}.
Here, we will explicitly show what is happening in the bulk, following the logic of \cite{Aalsma:2021bit}. For this we require the bulk dilaton solution \eqref{eq:backreactedbulkdilaton}.
At the Page time, which corresponds to setting $\alpha=\frac{1}{2}$ in \eqref{eq:backreactedbulkdilaton}, we discover the existence of a trapped region. In this two-dimensional setting the expansion scalars translate to $\partial_\pm \Phi$; trapped regions are therefore defined as $\partial_\pm \Phi < 0$, which translates to 

 \begin{equation}
    \begin{aligned}\label{eq:trappedregionconditions}
        x^+ \Big(c G( (x^+x^-)^2 - \ell^4) + 2x^+x^- \ell^2 \big(6\pi - c G \log \frac{x^+}{\ell}\big)\Big) &< 0~, \\
    x^+\Big( 6 \pi  - c G \log \frac{x^+}{\ell}\Big) &< 0~,
    \end{aligned}
    \end{equation}
respectively. These inequalities determine two curves bounding the trapped region. These curves are given by
    \begin{equation}\label{eq:trappedcurves}
    \begin{aligned}
      \gamma_1\,&:\quad  x^+ = \ell \exp\left(\frac{6\pi}{cG}\right)~, \\
      \gamma_2\,&:\quad  x^- = \frac{\ell^2}{cG x^+} \left( c G \log\frac{x^+}{\ell}  - 6\pi - \sqrt{\left(6\pi - c G \log\frac{x^+}{\ell} \right)^2 + (cG)^2} \right)~.
    \end{aligned}
    \end{equation}
Equality $\partial_+ \Phi = 0 = \partial_- \Phi$ is obtained at the pole, $(x^+, x^-) = (x^+_{\gamma_1}, - \ell^2/x^+_{\gamma_1})$. For a static observer at this pole, the trapped region occurs at the time
\begin{equation}\label{eq:trappedtime}
        t> t_\text{trapped} = \frac{6\pi\ell}{cG} = t_{\text{Page}}~,
    \end{equation}
with the analogous statement for the meta-observer in terms of $u$; see \figref{fig:trappedregion}. As can be seen explicitly in \figref{fig:trappedregion}, the trapped region prevents the radiation from reaching the static patch observer for times $t > t_\text{trapped}$. Moreover, the same is true for the meta-observer at future infinity, who does not have access to radiation for $u > u_\text{trapped}$. However, the Penrose-Hawking singularity theorem \cite{PhysRevLett.14.57,hawking_ellis_1973} does not immediately apply to the trapped region described here, as the NEC is violated. It can be shown by use of the quantum singularity theorem \cite{QuantumSingularityTheorem} that for $u > u_\text{trapped}$ a (quantum) singularity forms at future infinity \cite{Aalsma:2021bit}. 

\section{Discussion}
In this section we will summarise our findings on a more conceptual level and elaborate on open questions. We conclude by commenting on inflationary scenarios in this setup.

In this paper, we considered a three-dimensional de Sitter geometry in a setup which naturally supplies a segmentation into dynamical gravity on dS$_2$ and an auxiliary system located in the direction of the partial reduction. Since we are interested in modelling an evaporation process, we added conformal matter to our theory. Putting dynamical gravity in an out-of-equilibrium quantum state then indeed furnishes a decreasing Gibbons-Hawking entropy due to backreaction effects.
By analysing the behaviour of the backreacted dilaton solution, we discover two decoupling regions in which field theory descriptions arise. As should be expected, one of these regions is $\mathcal{I}^{+}$, in line with \cite{Stromingerdscft}. Perhaps more surprisingly, the second region is inside the static patch at the past cosmological horizon, which we interpret in the holographic `stretched horizon' picture of \cite{SusskindStretched, Susskind:2021omt}.
In both regions we can thus calculate the fine-grained entropy of the collected radiation via field theory considerations. 

\subsection{Quantum mechanics and the island}
Let us elaborate on this setup in the language of \cite{Raju:2020smc}.\footnote{See section 5.4 of \cite{Raju:2020smc}.} We will use the region at $\mathcal{I}^{+}$ to be explicit, but analogous statements hold for the static patch. The gravitational dynamics of the two-dimensional de Sitter space \eqref{eq:MilneDilaton} with coordinates $(\tau,\chi)$ reduce to that of a quantum mechanical boundary degree of freedom. Since the dilaton can be given a three-dimensional interpretation as an angle, this can be interpreted as transparency along a third direction $\varphi$, such that we are coupling the single quantum mechanical boundary degree of freedom to a matter CFT. We can now consider an imaginary interface in this non-gravitational theory, which factorises the Hilbert space into two subsystems: the quantum mechanical boundary degree of freedom coupled to a part of the bath, and the remainder of the bath. It is the entanglement entropy between these two systems that we are calculating.

For the calculation at future infinity we recover a naturally pure evaporating process. The static observer, more akin to an asymptotic AdS observer, requires a more involved argument to furnish a unitary process. Our results for the static observer are in line with \cite{Aalsma:2021bit}, in which a stretched horizon picture was advocated for this observer with the gravitational physics reducing to field theory considerations at the horizon. This is supported by our results. We can however even make a more general statement: at the past cosmological horizon we see a timelike realisation of the backreaction dynamics determined at $\mathcal{I}^{+}$. A natural question to ask is how far this connection can be pushed. Although we believe these comments are important for our understanding of various observers in de Sitter space, for this quantum state, the main takeaway should still be that for both observers we finally arrive at a tragedy as neither can recover any information. 

It is interesting to connect our construction to the more canonical island approach. For the BTZ construction of \cite{Verheijden} it seems the island can be identified with the region bounded by the RT surface of the thermal bath within the gravitating region, which lies outside of the horizon. While our approach naturally introduces a thermal bath, it is not immediately clear how to pinpoint the location of the island for the de Sitter case, as was done in \cite{Aalsma:2021bit}. It would be interesting to explore this further. To do so, it is also important to connect our results to \cite{CosmologyIslands}, in which three necessary criteria were constructed for the existence of an island. These implied that for pure de Sitter spacetimes there are no islands for an entangling region located at future infinity. To put our results into this context---and see how these conditions are evaded---, notice the following. First, we are considering a different quantum state: not the Hartle-Hawking/Bunch-Davies state but the Unruh-de Sitter state. Secondly, we find that the use of \eqref{eq:island} is only required for the static observer, whereas the no-go conjecture of \cite{CosmologyIslands} considers a region at future infinity. 

\subsection{Information recovery}
For evaporating black holes, small amounts of information thrown in after the Page time can be recovered from the Hawking radiation after waiting the so-called scrambling time $t\approx \frac{\beta}{2 \pi}\log S_{\text{BH}}$ \cite{Hayden:2007cs,Sekino:2008he,Shenker:2013pqa}. Does a similar story hold for the evaporating de Sitter horizon? Due to the occurrence of catastrophic backreaction at the Page time within our construction, this seems not to be the case. In \cite{MScThesisVerheijden, Aalsma:2021kle} a setup was presented in which recovery of information expelled through the cosmological horizon was analysed in the Bunch-Davies state using shock waves; and in \cite{Aalsma:2021kle} a concrete protocol was proposed for information transfer to the antipodal observer. The relevant timescale here, too, is the scrambling time. As at early times the Unruh-de Sitter state should be almost indistinguishable from the Bunch-Davies state, it might be expected that some form of information retrieval might be possible for (additional) information expelled through the cosmological horizon. It would be interesting to see if this expectation is indeed true and at which timescale this breaks down due to the deviation of the two quantum states.

\subsection{Inflationary perspective}
Finally, let us comment on inflationary physics. To do so, we use planar coordinates \eqref{eq:planarcoordinates}, and consider a scenario as explained in e.g.\ \cite{Chen:2020tes}. To the future of our de Sitter construction we glue flat space, corresponding to the old Universe in which gravitational effects can be neglected. This is also depicted in \figref{fig:trappedregion}. Future infinity constitutes the reheating surface, the transitory region between the inflating and the old Universe. This would give a simple model for analysing primordial fluctuations. Let us not consider the full setup but only make some tentative comments up to $\mathcal{I}^{+}$. The evaporating quantum state of \eqref{eq:UnruhforgeneralCFTglobal} expressed in terms of \eqref{eq:planarcoordinates} also leads to a net flux at future infinity. However, as the coordinates \eqref{eq:planarcoordinates} are ground state solutions of the Schwarzian theory, the conserved ADM quantity \eqref{eq:admquantity} vanishes on-shell. Contrast this with both the Milne coordinates \eqref{eq:Milne} and the global coordinates \eqref{eq:dilatonglobal} for which the ADM quantity \eqref{eq:admquantity} is related to the entropy of the cosmological horizon, such that the quantum state indeed captures the evaporating horizon. As already noted in \cite{backreactionFuture}, in the coordinates \eqref{eq:planarcoordinates} the stress tensor components are well-defined within the entire planar patch. Hence, for these coordinates the Unruh state is a natural alternative to the Bunch-Davies state. It would be interesting to pursue this direction further. It would also be interesting to connect \eqref{eq:AdMtoentropy} with the results of \cite{Balasubramanian:2001nb} and to understand if a first law may be constructed, linking a variation of the conserved quantity $K$ to a variation in the  entropy $S_{\text{dS},\alpha}$. 

In general, as inflation is a UV dependent problem, the island formula \eqref{eq:island} may play a pivotal role in understanding inflationary scenarios via fine-grained entropy considerations. As such it is important to understand in what way non-perturbative effects of the replica wormhole type are realised in inflationary models and how this changes semi-classical expectations. A fruitful avenue could be to consider potential entropy paradoxes for various subregions of the gravitionally prepared state depicted in \figref{fig:trappedregion}, as in \cite{Chen:2020tes}. In addition, for these inflationary setups it would be important to understand in how far higher-dimensional setups can evade the constraints of \cite{CosmologyIslands}.

\acknowledgments
It is a pleasure to thank Jan Pieter van der Schaar for useful discussions and comments on a draft. JKK would furthermore like to thank Alexandros Kanargias and Stefan F\"orste for helpful discussions. Work of EPV and EV is supported by the Spinoza grant and the Delta ITP consortium, a program of the NWO that is funded by the Dutch Ministry of Education, Culture and Science (OCW). Work of JKK was supported in part by the Heising-Simons Foundation, the Simons Foundation, and National Science Foundation Grant No. NSF PHY-1748958. JKK also gratefully acknowledges support by the Graduate School of Physics and Astronomy (BCGS).
JKK, EV and EPV would like to thank the Institute d'Études Scientifiques de Cargèse, where part of this work was conducted, for hospitality, and the attendees of the 2021 summer school `Old and New Frontiers in Quantum Field Theory' for a stimulating atmosphere. JKK finally would like to thank the University of Amsterdam for hospitality during the completion of this work. 
\vspace{1cm}

\begin{appendices}
\section{Coordinate systems}\label{app:coordinates}
In this appendix, we will for the sake of completeness give various two-dimensional coordinate systems for de Sitter space and the associated dilaton solutions. Note that the solutions we quote originate from dimensional reduction and not from an intrinsically two-dimensional setup. Concretely, this means that the dilaton solutions might differ from previous work on JT gravity in purely two-dimensional de Sitter. 

\paragraph{Global coordinates.}
In global coordinates $(T, \theta)$ the metric and dilaton are
    \begin{equation}
        ds^2 = - \dd T^2 + \ell^2 \cosh^2{\frac{T}{\ell}} \dd \theta^2~, \qquad \Phi = 2 \pi \alpha \sin\theta \cosh\frac{T}{\ell}~.
\end{equation}
Here, $\theta \in [0,\pi]$ and $T \in (-\infty, \infty )$. These coordinates are called global coordinates as they can be used to describe all of de Sitter space. 
 
\paragraph{Global conformal coordinates.}
In global conformal coordinates $(\sigma, \theta)$ the metric and dilaton are given as
    \begin{equation}\label{eq:globalconformal}
        ds^2 = \frac{\ell^2}{\cos^2\sigma} \lp - \dd \sigma^2 + \dd \theta^2 \rp~, \qquad \Phi = 2\pi \alpha \frac{\sin\theta}{\cos\sigma}~,
    \end{equation}
where $\sigma \in (-\frac{\pi}{2}, \frac{\pi}{2})$. These coordinates cover the full Penrose diagram of de Sitter, \figref{fig:Penrosediagram}. The transformation to go from global coordinates to global conformal coordinates is given by
    \begin{equation} \label{eq:globaltoglobalconf}
        \tan\frac{\sigma}{2} = \tanh \frac{T}{2\ell}~.
    \end{equation}
In terms of the Schwarzian equations of motion \eqref{eq:schwarzianeomtimeindependent}, \eqref{eq:globalconformal} corresponds to
\begin{align}\label{eq:globalx}
    x(u)&= 2 \ell \tan{\frac{u}{2 \ell}}\,.
\end{align}
\paragraph{Planar coordinates.}
In the flat slicing the metric and dilaton are given by
    \begin{equation}
        ds^2 = - \dd t^2 + e^{2t/\ell} \dd \rho^2~, \qquad \Phi = 2\pi \alpha \frac{\rho}{\ell} e^{t/\ell}~,
    \end{equation}
where $\rho \geq 0$ and $t \in (-\infty, \infty)$. These coordinates cover half of the de Sitter Penrose diagram (static patch + future patch). The transformation between planar and global coordinates is given by
    \begin{equation} \label{eq:planartoglobal}
    \begin{aligned}
        \rho &= \frac{\ell \cosh{\frac{T}{\ell}}\sin\theta}{\sinh \frac{T}{\ell} + \cos\theta \cosh\frac{T}{\ell}}~, \\
        e^{t/\ell} &= \sinh \frac{T}{\ell} + \cos\theta \cosh\frac{T}{\ell}~.
    \end{aligned}
    \end{equation}

\paragraph{Planar conformal coordinates.}
From the previous coordinate system, we can go to conformal time $\eta$ via
    \begin{equation} \label{eq:conftime}
        \eta = - \ell e^{-t/\ell}~,
    \end{equation}
where $\eta \leq 0$ with equality at $\mathcal{I}^+$. Then the metric and dilaton are given by
    \begin{equation}
        ds^2 = \frac{\ell^2}{\eta^2} \lp - \dd \eta^2 + \dd x^2 \rp~, \qquad \Phi = - 2\pi \alpha \frac{x}{\eta}~,
    \end{equation}
where we set $x = \rho$. These coordinates cover the same planar patch as the previous ones. 

Note that we can combine the coordinate transformations \eqref{eq:planartoglobal}, \eqref{eq:globaltoglobalconf} and \eqref{eq:conftime} to find a direct relation between planar conformal coordinates $(\eta, x)$ and global conformal coordinates $(\sigma, \theta)$: 
    \begin{equation}
    \begin{aligned}
        \eta &= - \frac{\ell \cos\sigma}{\cos\theta + \sin\sigma}~, \\
        x &= \frac{\ell \sin\theta}{\cos\theta + \sin\sigma}~.
    \end{aligned}
    \end{equation}

\paragraph{Kruskal coordinates.}
We can extend the planar coordinates to cover the entire Penrose diagram by defining Kruskal coordinates as follows:
    \begin{equation}
        x^+ = \eta + x~, \quad x^- = - \frac{\ell^2}{\eta - x}~.
    \end{equation}
Then the metric and dilaton are given by
    \begin{equation}
        ds^2 = - \frac{4\ell^4}{(\ell^2 - x^+x^-)^2} \dd x^+ \dd x^-~, \qquad \Phi = 2\pi \alpha \lp \frac{\ell^2 + x^+ x^-}{\ell^2 - x^+ x^-} \rp~.
    \end{equation}

\paragraph{Static patch coordinates.}
The static patch coordinates are defined with respect to the Kruskal coordinates as
    \begin{equation}
        x^+ = \ell e^{t/\ell} \sqrt{\frac{\ell - r}{\ell + r}}~, \quad x^- = - \ell e^{-t/\ell} \sqrt{\frac{\ell-r}{\ell+r}}~.
    \end{equation}
In terms of these coordinates, the metric and dilaton are
    \begin{equation}\label{eq:staticpatch2dcoordinates}
        ds^2 = - \lp 1 - \frac{r^2}{\ell^2} \rp \dd t^2 + \lp 1 - \frac{r^2}{\ell^2} \rp^{-1} \dd r^2~, \qquad \Phi = 2\pi \alpha \frac{r}{\ell}\,.
    \end{equation}
These coordinates only cover the static patch for an observer located at the south pole $r = 0$; their cosmological horizon is located at $r = \ell$.

We can also define the Kruskal coordinates via 
    \begin{equation}\label{eq:KruskaltostaticApp}
        x^{\pm}=\pm \ell e^{\pm \sigma^\pm/\ell} ~,
    \end{equation}
where we introduced null coordinates
    \begin{equation}\label{eq:staticnullcoordinates}
        \sigma^{\pm}=t \pm r^{*}\,.
    \end{equation}
Here, $r^{*}$ is a tortoise coordinate
    \begin{align}
        r^{*}=\int_0^{r} \frac{1}{f(r')} dr^{'} =\ell \,\text{arctanh}\left( \frac{r}{\ell} \right)\,,
    \end{align}
which only holds for $r < \ell$ and hence only covers the static patch. Note that the south pole ($r=0$) is at $r^{*}=0$ and the cosmological horizon ($r=\ell$) at $r^{*}=\infty$. In terms of the null coordinates $\sigma^\pm$ we recover the metric \eqref{eq:staticpatchnullcoordinates}. Note that \eqref{eq:KruskaltostaticApp} is equivalent to the coordinate change between Rindler and Minkowski.

\paragraph{Milne coordinates.}
Finally, from the static patch coordinates we can analytically continue across the future horizon to describe the future or Milne patch:\footnote{Note that this continuation differs slightly from the one presented in \cite{MaldacenaTuriaci}. We believe the one given here is correct.} 
    \begin{equation}\label{eq:analyticcontinuation}
        \tau = i \ell \arccos \frac{r}{\ell}~, \quad \chi = t~,
    \end{equation}
which gives
    \begin{equation} \label{eq:MilneDilaton}
        ds^2 = - \dd \tau^2 + \sinh^2 \frac{\tau}{\ell} \dd \chi^2\,,~ \quad \quad \Phi=2 \pi \alpha \cosh{\frac{\tau}{\ell}}.
    \end{equation}
Unlike the static patch solution, the Milne solution does not exhibit time translation symmetry. However, they are essentially the same solutions connected by analytic continuation; in the two patches, the isometries are actualised in a different manner. 

\end{appendices}

\newpage
\bibliographystyle{JHEP}
\bibliography{library}

\providecommand{\href}[2]{#2}\begingroup\raggedright\begin{thebibliography}{10}

\bibitem{GibbonsHawking}
G.W.~Gibbons and S.W.~Hawking, \emph{{Cosmological Event Horizons,
  Thermodynamics, and Particle Creation}},
  \href{https://doi.org/10.1103/PhysRevD.15.2738}{\emph{Phys. Rev. D}
  {\bfseries 15} (1977) 2738}.

\bibitem{Wittendesitter}
E.~Witten, \emph{{Quantum gravity in de Sitter space}},  in \emph{{Strings
  2001: International Conference}}, 6, 2001
  [\href{https://arxiv.org/abs/hep-th/0106109}{{\ttfamily hep-th/0106109}}].

\bibitem{troublewithds}
N.~Goheer, M.~Kleban and L.~Susskind, \emph{{The Trouble with de Sitter
  space}}, \href{https://doi.org/10.1088/1126-6708/2003/07/056}{\emph{JHEP}
  {\bfseries 07} (2003) 056}
  [\href{https://arxiv.org/abs/hep-th/0212209}{{\ttfamily hep-th/0212209}}].

\bibitem{ParikhErik}
M.K.~Parikh and E.P.~Verlinde, \emph{{De Sitter holography with a finite number
  of states}}, \href{https://doi.org/10.1088/1126-6708/2005/01/054}{\emph{JHEP}
  {\bfseries 01} (2005) 054}
  [\href{https://arxiv.org/abs/hep-th/0410227}{{\ttfamily hep-th/0410227}}].

\bibitem{Erik2}
M.K.~Parikh, I.~Savonije and E.P.~Verlinde, \emph{{Elliptic de Sitter space:
  dS/Z(2)}}, \href{https://doi.org/10.1103/PhysRevD.67.064005}{\emph{Phys. Rev.
  D} {\bfseries 67} (2003) 064005}
  [\href{https://arxiv.org/abs/hep-th/0209120}{{\ttfamily hep-th/0209120}}].

\bibitem{higherspinds}
D.~Anninos, F.~Denef, R.~Monten and Z.~Sun, \emph{{Higher Spin de Sitter
  Hilbert Space}}, \href{https://doi.org/10.1007/JHEP10(2019)071}{\emph{JHEP}
  {\bfseries 10} (2019) 071}
  [\href{https://arxiv.org/abs/1711.10037}{{\ttfamily 1711.10037}}].

\bibitem{Dong:2018cuv}
X.~Dong, E.~Silverstein and G.~Torroba, \emph{{De Sitter Holography and
  Entanglement Entropy}},
  \href{https://doi.org/10.1007/JHEP07(2018)050}{\emph{JHEP} {\bfseries 07}
  (2018) 050} [\href{https://arxiv.org/abs/1804.08623}{{\ttfamily
  1804.08623}}].

\bibitem{Geng:2019bnn}
H.~Geng, S.~Grieninger and A.~Karch, \emph{{Entropy, Entanglement and Swampland
  Bounds in DS/dS}}, \href{https://doi.org/10.1007/JHEP06(2019)105}{\emph{JHEP}
  {\bfseries 06} (2019) 105}
  [\href{https://arxiv.org/abs/1904.02170}{{\ttfamily 1904.02170}}].

\bibitem{RyuTakayanagi}
S.~Ryu and T.~Takayanagi, \emph{{Holographic derivation of entanglement entropy
  from AdS/CFT}},
  \href{https://doi.org/10.1103/PhysRevLett.96.181602}{\emph{Phys. Rev. Lett.}
  {\bfseries 96} (2006) 181602}
  [\href{https://arxiv.org/abs/hep-th/0603001}{{\ttfamily hep-th/0603001}}].

\bibitem{HubenyRT}
V.E.~Hubeny, M.~Rangamani and T.~Takayanagi, \emph{{A Covariant holographic
  entanglement entropy proposal}},
  \href{https://doi.org/10.1088/1126-6708/2007/07/062}{\emph{JHEP} {\bfseries
  07} (2007) 062} [\href{https://arxiv.org/abs/0705.0016}{{\ttfamily
  0705.0016}}].

\bibitem{EngelhardtQES}
N.~Engelhardt and A.C.~Wall, \emph{{Quantum Extremal Surfaces: Holographic
  Entanglement Entropy beyond the Classical Regime}},
  \href{https://doi.org/10.1007/JHEP01(2015)073}{\emph{JHEP} {\bfseries 01}
  (2015) 073} [\href{https://arxiv.org/abs/1408.3203}{{\ttfamily 1408.3203}}].

\bibitem{Faulkner:2013ana}
T.~Faulkner, A.~Lewkowycz and J.~Maldacena, \emph{{Quantum corrections to
  holographic entanglement entropy}},
  \href{https://doi.org/10.1007/JHEP11(2013)074}{\emph{JHEP} {\bfseries 11}
  (2013) 074} [\href{https://arxiv.org/abs/1307.2892}{{\ttfamily 1307.2892}}].

\bibitem{Barrella:2013wja}
T.~Barrella, X.~Dong, S.A.~Hartnoll and V.L.~Martin, \emph{{Holographic
  entanglement beyond classical gravity}},
  \href{https://doi.org/10.1007/JHEP09(2013)109}{\emph{JHEP} {\bfseries 09}
  (2013) 109} [\href{https://arxiv.org/abs/1306.4682}{{\ttfamily 1306.4682}}].

\bibitem{Penington1}
G.~Penington, \emph{{Entanglement Wedge Reconstruction and the Information
  Paradox}}, \href{https://doi.org/10.1007/JHEP09(2020)002}{\emph{JHEP}
  {\bfseries 09} (2020) 002}
  [\href{https://arxiv.org/abs/1905.08255}{{\ttfamily 1905.08255}}].

\bibitem{AlmheiriEMM}
A.~Almheiri, N.~Engelhardt, D.~Marolf and H.~Maxfield, \emph{{The entropy of
  bulk quantum fields and the entanglement wedge of an evaporating black
  hole}}, \href{https://doi.org/10.1007/JHEP12(2019)063}{\emph{JHEP} {\bfseries
  12} (2019) 063} [\href{https://arxiv.org/abs/1905.08762}{{\ttfamily
  1905.08762}}].

\bibitem{AlmheiriMMZ}
A.~Almheiri, R.~Mahajan, J.~Maldacena and Y.~Zhao, \emph{{The Page curve of
  Hawking radiation from semiclassical geometry}},
  \href{https://doi.org/10.1007/JHEP03(2020)149}{\emph{JHEP} {\bfseries 03}
  (2020) 149} [\href{https://arxiv.org/abs/1908.10996}{{\ttfamily
  1908.10996}}].

\bibitem{Page1}
D.N.~Page, \emph{{Information in black hole radiation}},
  \href{https://doi.org/10.1103/PhysRevLett.71.3743}{\emph{Phys. Rev. Lett.}
  {\bfseries 71} (1993) 3743}
  [\href{https://arxiv.org/abs/hep-th/9306083}{{\ttfamily hep-th/9306083}}].

\bibitem{Page2}
D.N.~Page, \emph{{Time Dependence of Hawking Radiation Entropy}},
  \href{https://doi.org/10.1088/1475-7516/2013/09/028}{\emph{JCAP} {\bfseries
  1309} (2013) 028} [\href{https://arxiv.org/abs/1301.4995}{{\ttfamily
  1301.4995}}].

\bibitem{Raju1}
H.~Geng, A.~Karch, C.~Perez-Pardavila, S.~Raju, L.~Randall, M.~Riojas et~al.,
  \emph{{Information Transfer with a Gravitating Bath}},
  \href{https://doi.org/10.21468/SciPostPhys.10.5.103}{\emph{SciPost Phys.}
  {\bfseries 10} (2021) 103}
  [\href{https://arxiv.org/abs/2012.04671}{{\ttfamily 2012.04671}}].

\bibitem{Raju2}
A.~Laddha, S.G.~Prabhu, S.~Raju and P.~Shrivastava, \emph{{The Holographic
  Nature of Null Infinity}},
  \href{https://doi.org/10.21468/SciPostPhys.10.2.041}{\emph{SciPost Phys.}
  {\bfseries 10} (2021) 041}
  [\href{https://arxiv.org/abs/2002.02448}{{\ttfamily 2002.02448}}].

\bibitem{AlmheiriOutside}
A.~Almheiri, R.~Mahajan and J.~Maldacena, \emph{{Islands outside the horizon}},
   \href{https://arxiv.org/abs/1910.11077}{{\ttfamily 1910.11077}}.

\bibitem{Hollowood1}
T.J.~Hollowood and S.P.~Kumar, \emph{{Islands and Page Curves for Evaporating
  Black Holes in JT Gravity}},
  \href{https://doi.org/10.1007/JHEP08(2020)094}{\emph{JHEP} {\bfseries 08}
  (2020) 094} [\href{https://arxiv.org/abs/2004.14944}{{\ttfamily
  2004.14944}}].

\bibitem{MyersEquil}
H.Z.~Chen, Z.~Fisher, J.~Hernandez, R.C.~Myers and S.-M.~Ruan,
  \emph{{Evaporating Black Holes Coupled to a Thermal Bath}},
  \href{https://doi.org/10.1007/JHEP01(2021)065}{\emph{JHEP} {\bfseries 01}
  (2021) 065} [\href{https://arxiv.org/abs/2007.11658}{{\ttfamily
  2007.11658}}].

\bibitem{Almheirihigherdimensions}
A.~Almheiri, R.~Mahajan and J.E.~Santos, \emph{{Entanglement islands in higher
  dimensions}},
  \href{https://doi.org/10.21468/SciPostPhys.9.1.001}{\emph{SciPost Phys.}
  {\bfseries 9} (2020) 001} [\href{https://arxiv.org/abs/1911.09666}{{\ttfamily
  1911.09666}}].

\bibitem{Gautason:2020tmk}
F.F.~Gautason, L.~Schneiderbauer, W.~Sybesma and L.~Thorlacius, \emph{{Page
  Curve for an Evaporating Black Hole}},
  \href{https://doi.org/10.1007/JHEP05(2020)091}{\emph{JHEP} {\bfseries 05}
  (2020) 091} [\href{https://arxiv.org/abs/2004.00598}{{\ttfamily
  2004.00598}}].

\bibitem{Anegawa:2020ezn}
T.~Anegawa and N.~Iizuka, \emph{{Notes on islands in asymptotically flat 2d
  dilaton black holes}},
  \href{https://doi.org/10.1007/JHEP07(2020)036}{\emph{JHEP} {\bfseries 07}
  (2020) 036} [\href{https://arxiv.org/abs/2004.01601}{{\ttfamily
  2004.01601}}].

\bibitem{Hashimoto:2020cas}
K.~Hashimoto, N.~Iizuka and Y.~Matsuo, \emph{{Islands in Schwarzschild black
  holes}}, \href{https://doi.org/10.1007/JHEP06(2020)085}{\emph{JHEP}
  {\bfseries 06} (2020) 085}
  [\href{https://arxiv.org/abs/2004.05863}{{\ttfamily 2004.05863}}].

\bibitem{Hartman:2020swn}
T.~Hartman, E.~Shaghoulian and A.~Strominger, \emph{{Islands in Asymptotically
  Flat 2D Gravity}}, \href{https://doi.org/10.1007/JHEP07(2020)022}{\emph{JHEP}
  {\bfseries 07} (2020) 022}
  [\href{https://arxiv.org/abs/2004.13857}{{\ttfamily 2004.13857}}].

\bibitem{higherD2}
H.~Geng and A.~Karch, \emph{{Massive islands}},
  \href{https://doi.org/10.1007/JHEP09(2020)121}{\emph{JHEP} {\bfseries 09}
  (2020) 121} [\href{https://arxiv.org/abs/2006.02438}{{\ttfamily
  2006.02438}}].

\bibitem{Alishahiha:2020qza}
M.~Alishahiha, A.~Faraji~Astaneh and A.~Naseh, \emph{{Island in the presence of
  higher derivative terms}},
  \href{https://doi.org/10.1007/JHEP02(2021)035}{\emph{JHEP} {\bfseries 02}
  (2021) 035} [\href{https://arxiv.org/abs/2005.08715}{{\ttfamily
  2005.08715}}].

\bibitem{Verheijden}
E.~Verheijden and E.~Verlinde, \emph{{From the BTZ black hole to JT gravity:
  geometrizing the island}},
  \href{https://doi.org/10.1007/JHEP11(2021)092}{\emph{JHEP} {\bfseries 11}
  (2021) 092} [\href{https://arxiv.org/abs/2102.00922}{{\ttfamily
  2102.00922}}].

\bibitem{Maloney}
J.~Cotler, K.~Jensen and A.~Maloney, \emph{{Low-dimensional de Sitter quantum
  gravity}}, \href{https://doi.org/10.1007/JHEP06(2020)048}{\emph{JHEP}
  {\bfseries 06} (2020) 048}
  [\href{https://arxiv.org/abs/1905.03780}{{\ttfamily 1905.03780}}].

\bibitem{MaldacenaTuriaci}
J.~Maldacena, G.J.~Turiaci and Z.~Yang, \emph{{Two dimensional Nearly de Sitter
  gravity}}, \href{https://doi.org/10.1007/JHEP01(2021)139}{\emph{JHEP}
  {\bfseries 01} (2021) 139}
  [\href{https://arxiv.org/abs/1904.01911}{{\ttfamily 1904.01911}}].

\bibitem{Blommaert:2020tht}
A.~Blommaert, \emph{{Searching for butterflies in dS JT gravity}},
  \href{https://arxiv.org/abs/2010.14539}{{\ttfamily 2010.14539}}.

\bibitem{Sybesma}
W.~Sybesma, \emph{{Pure de Sitter space and the island moving back in time}},
  \href{https://doi.org/10.1088/1361-6382/abff9a}{\emph{Class. Quant. Grav.}
  {\bfseries 38} (2021) 145012}
  [\href{https://arxiv.org/abs/2008.07994}{{\ttfamily 2008.07994}}].

\bibitem{BalasubramanianDSislands}
V.~Balasubramanian, A.~Kar and T.~Ugajin, \emph{{Islands in de Sitter space}},
  \href{https://doi.org/10.1007/JHEP02(2021)072}{\emph{JHEP} {\bfseries 02}
  (2021) 072} [\href{https://arxiv.org/abs/2008.05275}{{\ttfamily
  2008.05275}}].

\bibitem{CosmologyIslands}
T.~Hartman, Y.~Jiang and E.~Shaghoulian, \emph{{Islands in cosmology}},
  \href{https://doi.org/10.1007/JHEP11(2020)111}{\emph{JHEP} {\bfseries 11}
  (2020) 111} [\href{https://arxiv.org/abs/2008.01022}{{\ttfamily
  2008.01022}}].

\bibitem{Geng:2021wcq}
H.~Geng, Y.~Nomura and H.-Y.~Sun, \emph{{Information paradox and its resolution
  in de Sitter holography}},
  \href{https://doi.org/10.1103/PhysRevD.103.126004}{\emph{Phys. Rev. D}
  {\bfseries 103} (2021) 126004}
  [\href{https://arxiv.org/abs/2103.07477}{{\ttfamily 2103.07477}}].

\bibitem{Aalsma:2021bit}
L.~Aalsma and W.~Sybesma, \emph{{The Price of Curiosity: Information Recovery
  in de Sitter Space}},
  \href{https://doi.org/10.1007/JHEP05(2021)291}{\emph{JHEP} {\bfseries 05}
  (2021) 291} [\href{https://arxiv.org/abs/2104.00006}{{\ttfamily
  2104.00006}}].

\bibitem{backreactionFuture}
L.~Aalsma, M.~Parikh and J.P.~Van Der~Schaar, \emph{{Back(reaction) to the
  Future in the Unruh-de Sitter State}},
  \href{https://doi.org/10.1007/JHEP11(2019)136}{\emph{JHEP} {\bfseries 11}
  (2019) 136} [\href{https://arxiv.org/abs/1905.02714}{{\ttfamily
  1905.02714}}].

\bibitem{PhysRevD.28.2960}
J.B.~Hartle and S.W.~Hawking, \emph{Wave function of the universe},
  \href{https://doi.org/10.1103/PhysRevD.28.2960}{\emph{Phys. Rev. D}
  {\bfseries 28} (1983) 2960}.

\bibitem{Danielsson:2002qh}
U.H.~Danielsson, \emph{{Inflation, holography, and the choice of vacuum in de
  Sitter space}},
  \href{https://doi.org/10.1088/1126-6708/2002/07/040}{\emph{JHEP} {\bfseries
  07} (2002) 040} [\href{https://arxiv.org/abs/hep-th/0205227}{{\ttfamily
  hep-th/0205227}}].

\bibitem{AlmheiriPolchinski}
A.~Almheiri and J.~Polchinski, \emph{{Models of AdS$_{2}$ backreaction and
  holography}}, \href{https://doi.org/10.1007/JHEP11(2015)014}{\emph{JHEP}
  {\bfseries 11} (2015) 014} [\href{https://arxiv.org/abs/1402.6334}{{\ttfamily
  1402.6334}}].

\bibitem{MaldacenaNAdS}
J.~Maldacena, D.~Stanford and Z.~Yang, \emph{{Conformal symmetry and its
  breaking in two dimensional Nearly Anti-de-Sitter space}},
  \href{https://doi.org/10.1093/ptep/ptw124}{\emph{PTEP} {\bfseries 2016}
  (2016) 12C104} [\href{https://arxiv.org/abs/1606.01857}{{\ttfamily
  1606.01857}}].

\bibitem{HermanBackreaction}
J.~Engelsöy, T.G.~Mertens and H.~Verlinde, \emph{{An investigation of
  AdS$_{2}$ backreaction and holography}},
  \href{https://doi.org/10.1007/JHEP07(2016)139}{\emph{JHEP} {\bfseries 07}
  (2016) 139} [\href{https://arxiv.org/abs/1606.03438}{{\ttfamily
  1606.03438}}].

\bibitem{PhysRevD.15.2088}
S.M.~Christensen and S.A.~Fulling, \emph{Trace anomalies and the hawking
  effect}, \href{https://doi.org/10.1103/PhysRevD.15.2088}{\emph{Phys. Rev. D}
  {\bfseries 15} (1977) 2088}.

\bibitem{Lohiya_1978}
D.~Lohiya, \emph{Trace anomalies in a two-dimensional de sitter metric and
  black-body radiation},
  \href{https://doi.org/10.1088/0305-4470/11/7/020}{\emph{Journal of Physics A:
  Mathematical and General} {\bfseries 11} (1978) 1335}.

\bibitem{Hawkingarealaw}
S.W.~Hawking, \emph{Gravitational radiation from colliding black holes},
  \href{https://doi.org/10.1103/PhysRevLett.26.1344}{\emph{Phys. Rev. Lett.}
  {\bfseries 26} (1971) 1344}.

\bibitem{Bousso:2015mqa}
R.~Bousso and N.~Engelhardt, \emph{{New Area Law in General Relativity}},
  \href{https://doi.org/10.1103/PhysRevLett.115.081301}{\emph{Phys. Rev. Lett.}
  {\bfseries 115} (2015) 081301}
  [\href{https://arxiv.org/abs/1504.07627}{{\ttfamily 1504.07627}}].

\bibitem{Maldacena:2001kr}
J.M.~Maldacena, \emph{{Eternal black holes in anti-de Sitter}},
  \href{https://doi.org/10.1088/1126-6708/2003/04/021}{\emph{JHEP} {\bfseries
  04} (2003) 021} [\href{https://arxiv.org/abs/hep-th/0106112}{{\ttfamily
  hep-th/0106112}}].

\bibitem{Parikh:2004wh}
M.K.~Parikh and E.P.~Verlinde, \emph{{De Sitter holography with a finite number
  of states}}, \href{https://doi.org/10.1088/1126-6708/2005/01/054}{\emph{JHEP}
  {\bfseries 01} (2005) 054}
  [\href{https://arxiv.org/abs/hep-th/0410227}{{\ttfamily hep-th/0410227}}].

\bibitem{Raju:2021lwh}
S.~Raju, \emph{{Failure of the split property in gravity and the information
  paradox}},  \href{https://arxiv.org/abs/2110.05470}{{\ttfamily 2110.05470}}.

\bibitem{Susskind:2021omt}
L.~Susskind, \emph{{De Sitter Holography: Fluctuations, Anomalous Symmetry, and
  Wormholes}}, \href{https://doi.org/10.3390/universe7120464}{\emph{Universe}
  {\bfseries 7} (2021) 464} [\href{https://arxiv.org/abs/2106.03964}{{\ttfamily
  2106.03964}}].

\bibitem{SusskindStretched}
L.~Susskind, L.~Thorlacius and J.~Uglum, \emph{{The Stretched horizon and black
  hole complementarity}},
  \href{https://doi.org/10.1103/PhysRevD.48.3743}{\emph{Phys. Rev. D}
  {\bfseries 48} (1993) 3743}
  [\href{https://arxiv.org/abs/hep-th/9306069}{{\ttfamily hep-th/9306069}}].

\bibitem{Stromingerdscft}
A.~Strominger, \emph{{The dS / CFT correspondence}},
  \href{https://doi.org/10.1088/1126-6708/2001/10/034}{\emph{JHEP} {\bfseries
  10} (2001) 034} [\href{https://arxiv.org/abs/hep-th/0106113}{{\ttfamily
  hep-th/0106113}}].

\bibitem{Calabrese:2004eu}
P.~Calabrese and J.L.~Cardy, \emph{{Entanglement entropy and quantum field
  theory}}, \href{https://doi.org/10.1088/1742-5468/2004/06/P06002}{\emph{J.
  Stat. Mech.} {\bfseries 0406} (2004) P06002}
  [\href{https://arxiv.org/abs/hep-th/0405152}{{\ttfamily hep-th/0405152}}].

\bibitem{OoguriBao}
N.~Bao and H.~Ooguri, \emph{{Distinguishability of black hole microstates}},
  \href{https://doi.org/10.1103/PhysRevD.96.066017}{\emph{Phys. Rev. D}
  {\bfseries 96} (2017) 066017}
  [\href{https://arxiv.org/abs/1705.07943}{{\ttfamily 1705.07943}}].

\bibitem{PhysRevLett.14.57}
R.~Penrose, \emph{Gravitational collapse and space-time singularities},
  \href{https://doi.org/10.1103/PhysRevLett.14.57}{\emph{Phys. Rev. Lett.}
  {\bfseries 14} (1965) 57}.

\bibitem{hawking_ellis_1973}
S.W.~Hawking and G.F.R.~Ellis, \emph{The Large Scale Structure of Space-Time},
  Cambridge Monographs on Mathematical Physics, Cambridge University Press
  (1973),
  \href{https://doi.org/10.1017/CBO9780511524646}{10.1017/CBO9780511524646}.

\bibitem{QuantumSingularityTheorem}
A.C.~Wall, \emph{{The Generalized Second Law implies a Quantum Singularity
  Theorem}}, \href{https://doi.org/10.1088/0264-9381/30/19/199501}{\emph{Class.
  Quant. Grav.} {\bfseries 30} (2013) 165003}
  [\href{https://arxiv.org/abs/1010.5513}{{\ttfamily 1010.5513}}].

\bibitem{Raju:2020smc}
S.~Raju, \emph{{Lessons from the information paradox}},
  \href{https://doi.org/10.1016/j.physrep.2021.10.001}{\emph{Phys. Rept.}
  {\bfseries 943} (2022) 2187}
  [\href{https://arxiv.org/abs/2012.05770}{{\ttfamily 2012.05770}}].

\bibitem{Hayden:2007cs}
P.~Hayden and J.~Preskill, \emph{{Black holes as mirrors: Quantum information
  in random subsystems}},
  \href{https://doi.org/10.1088/1126-6708/2007/09/120}{\emph{JHEP} {\bfseries
  09} (2007) 120} [\href{https://arxiv.org/abs/0708.4025}{{\ttfamily
  0708.4025}}].

\bibitem{Sekino:2008he}
Y.~Sekino and L.~Susskind, \emph{{Fast Scramblers}},
  \href{https://doi.org/10.1088/1126-6708/2008/10/065}{\emph{JHEP} {\bfseries
  10} (2008) 065} [\href{https://arxiv.org/abs/0808.2096}{{\ttfamily
  0808.2096}}].

\bibitem{Shenker:2013pqa}
S.H.~Shenker and D.~Stanford, \emph{{Black holes and the butterfly effect}},
  \href{https://doi.org/10.1007/JHEP03(2014)067}{\emph{JHEP} {\bfseries 03}
  (2014) 067} [\href{https://arxiv.org/abs/1306.0622}{{\ttfamily 1306.0622}}].

\bibitem{MScThesisVerheijden}
E.~Verheijden, \emph{{Traversable wormholes, shock waves, and de Sitter
  space}},  Master's thesis, University of Amsterdam, 2018.

\bibitem{Aalsma:2021kle}
L.~Aalsma, A.~Cole, E.~Morvan, J.P.~van~der Schaar and G.~Shiu, \emph{{Shocks
  and information exchange in de Sitter space}},
  \href{https://doi.org/10.1007/JHEP10(2021)104}{\emph{JHEP} {\bfseries 10}
  (2021) 104} [\href{https://arxiv.org/abs/2105.12737}{{\ttfamily
  2105.12737}}].

\bibitem{Chen:2020tes}
Y.~Chen, V.~Gorbenko and J.~Maldacena, \emph{{Bra-ket wormholes in
  gravitationally prepared states}},
  \href{https://doi.org/10.1007/JHEP02(2021)009}{\emph{JHEP} {\bfseries 02}
  (2021) 009} [\href{https://arxiv.org/abs/2007.16091}{{\ttfamily
  2007.16091}}].

\bibitem{Balasubramanian:2001nb}
V.~Balasubramanian, J.~de~Boer and D.~Minic, \emph{{Mass, entropy and
  holography in asymptotically de Sitter spaces}},
  \href{https://doi.org/10.1103/PhysRevD.65.123508}{\emph{Phys. Rev. D}
  {\bfseries 65} (2002) 123508}
  [\href{https://arxiv.org/abs/hep-th/0110108}{{\ttfamily hep-th/0110108}}].

\end{thebibliography}\endgroup

\end{document}